\documentclass[aip,pop,reprint,notitlepage,nofootinbib]{revtex4-1}

\usepackage{anysize}

\usepackage[utf8,koi8-r]{inputenc}
\usepackage[T2C]{fontenc}
\usepackage{mathtext}

\usepackage{amsmath}
\usepackage{amssymb}
\usepackage{bm}
\usepackage{hyperref}
\usepackage{graphics,graphicx}
\usepackage{multirow}

\usepackage{array}   
\newcolumntype{L}{>{$}l<{$}} 
\newcolumntype{C}{>{$}c<{$}} 
\newcolumntype{R}{>{$}r<{$}} 

\newcommand{\gn}{\bar{g}}
\newcommand{\xn}{\bar{X}}

\renewcommand{\i}{\alpha}
\renewcommand{\j}{\beta}

\marginsize{0.63in}{0.63in}{0.45in}{0.45in}

\begin{document}
\begin{abstract}
The derivation of the multi-temperature generalized Zhdanov closure is provided starting from the most general form of the left hand side of the moment averaged kinetic equation with the Sonine-Hermite polynomial ansatz for an arbitrary number of moments. The process of arriving at the reduced higher-order moment equations, with its assumptions and approximations, is explicitly outlined. The generalized multi-species, multi-temperature coefficients from the authors' previous article are used to compute values of higher order moments such as heat flux in terms of the lower order moments. Transport coefficients and the friction and thermal forces for magnetic confinement fusion relevant cases with the generalized coefficients are compared to the scheme with the single-temperature coefficients previously provided by Zhdanov et al. It is found that the $21N$-moment multi-temperature coefficients are adequate for most cases relevant to fusion. Furthermore, the $21N$-moment scheme is also tested against the trace approximation to determine the range of validity of the trace approximation with respect to fusion relevant plasmas. Possible refinements to the closure scheme are illustrated as well, in order to account for quantities which might be significant in certain schemes such as the drift approximation.  
\end{abstract} 
 
\title{Multi-temperature Generalized Zhdanov Closure for Scrape-Off Layer/Edge
Applications}
\author{M.\,Raghunathan}
\author{Y.\,Marandet}
\affiliation{Aix-Marseille Univ., CNRS, PIIM, Marseille, France}
\author{H.\,Bufferand} 
\author{G.\,Ciraolo}
\author{Ph.\,Ghendrih}
\author{P.\,Tamain}
\affiliation{IRFM-CEA, F-13108 Saint-Paul-Lez-Durance, France}
\author{E.\,Serre}
\affiliation{Aix-Marseille Univ., CNRS, M2P2, Marseille, France}
\maketitle

\section{Introduction}

Study of plasma transport in presence of impurities in the SOL/edge of current and next-step fusion devices such as WEST, JET and ITER, remains a key topic in magnetic confinement fusion research. Generally, SOL/edge plasmas are treated in the fluid domain by numerical fluid codes usually coupled to kinetic neutrals. The modelling of the friction and thermal forces in such codes is of paramount importance, especially for impurities, as the presence and transport of impurities in the SOL/edge of tokamaks largely depends on the competition between the friction and thermal forces. Generally, for a simple ion-electron plasma of a fixed charge state, there are low-moment closure schemes already available, such as by Braginskii\cite{braginskii_transport_1965}, which help compute these friction and thermal forces using the Landau collision operator. Such a scheme has also been generalized to a higher number of moments in the past\cite{kaneko_transport_1960,kaneko_electrical_1978,kaneko_thermal_1980} for the case of a single-temperature plasma using the linearized Boltzmann operator, and more recently for a multi-temperature plasma using the Landau operator\cite{ji_closure_2013}. These remarkable works have used a large number of moments to achieve convergence of transport coefficients to an increasing degree. However, firstly, there does not seem to be such a work present for the case of multi-temperature plasmas using the Boltzmann operator. Furthermore, the ion-electron case presents a few opportunities to simplify the calculations that cannot be straightforwardly extended to the case of a plasma with multiple species. The aforementioned scheme also solves the reduced set of equations for stationery solutions of plasmadynamical (such as density, momentum density and energy density) and other higher-order thermodynamic quantities (such as heat-flux and the stress tensor). In fluid simulations however, the plasmadynamical quantities are simulated by a time evolution scheme and are not subject to the aforementioned reductions. Therefore, one needs a solution that seeks a relatively stationary state for the higher-order thermodynamic quantities in terms of the variable plasmadynamical quantities. 

In order to remedy these issues, a closure scheme was suggested for the higher-order moments in the linear transport regime\cite{balescu_transport_1988}, derived from the linearized Boltzmann equation, by Zhdanov et al\cite{zhdanov_transport_2002,yushmanov_diffusion_1980,zhdanov_pfirschschlueter_1977}. The closure scheme, combined with their specific method of obtaining solutions for the higher-order moments such as heat-flux and the stress tensor, assuming collisions at common temperature, has come to be known as the Zhdanov closure.  The Zhdanov closure has been implemented in a variety of numerical SOL/edge fluid codes, such as Soledge3x-EIRENE\cite{bufferand_2019}, B2-EIRENE\cite{fichtmuller_multi-species_1998}, SOLPS\cite{sytova_derivation_2020,makarov_2021}, and EDGE2D\cite{bergmann_implementation_1996}.

Towards this, in the previous article by the authors\cite{raghunathan_generalized_2021}, we derived collision coefficients from the moment-averaged linearized Boltzmann collision operator for a multi-temperature case. The collision coefficients expressed themselves then as linear combinations of product of  terms depending on the mass and temperature ratios, and terms depending on the potential of interaction among the colliding particles. 
The coefficients were then compared to each other, in addition to a single-temperature set provided by Zhdanov et al\cite{yushmanov_diffusion_1980,zhdanov_pfirschschlueter_1977,zhdanov_transport_2002}, and a range of validity was provided for the the single-temperature coefficients. We also illustrated, using the obtained coefficients, a $13N$-moment closure scheme in the linear transport regime, showing some of the assumptions behind such a scheme, while establishing the effects of the different sets of coefficients on computed physical quantities such as viscosity and the augmentation in the friction force. We concluded, on the basis of these, that in case of large temperature differences, any set of multi-temperature collision coefficients was more reliable than the single-temperature set.
Furthermore, it is generally argued\cite{zhdanov_transport_2002,balescu_transport_1988}, that $13N$-moments are not adequate for correctly obtaining the values of the transport coefficients. Thus, the question remained open, in the scope of that article, as to whether going beyond the $21N$-moment scheme would help. Furthermore, some assumptions behind applying the linear transport regime to higher-order moments were also not explicitly stated. This article seeks to address these topics.

Furthermore, in the past, there have been some attempts at understanding the forces on impurities considering them to be in the trace approximation\cite{braginskii_transport_1965,chapman_thermal_1958,rutherford_impurity_1974,balescu_transport_1988}. Some code packages also used to have some trace limit schemes for treating impurities before implementing the $21N$-moment single-temperature Zhdanov closure\cite{bufferand_near_2013,rozhansky_momentum_2015}. Thus, it is also of interest to study the transport coefficients and the thermal and friction forces for the trace approximation against the Zhdanov closure, and thus provide a range of validity for the trace approximation scheme. 

In this article, in Sec.\,\ref{sec:lhs}, we introduce the Sonine-Hermite polynomials used, and first re-derive the most general moment-averaged fluid left hand side (LHS) term for an arbitrary number of moments. In Sec.\,\ref{sec:closure}, we illustrate, step-by-step, the various approximations that reduce the obtained parallel and longitudinal fluid equations to linear forms that can be inverted in order to obtain values for key quantities such as the heat-flux and the stress-tensor, and the friction and thermal forces. We then test the convergence of the scheme with increasing number of moments in \ref{sec:tests}, comparing the transport coefficients such as the longitudinal viscosity and parallel thermal conductivities, and the convergence of the friction and thermal forces. These tests are performed for a choice of four fusion relevant cases viz. deuterium-tritium as a light multi-species plasma, Carbon, Argon as mid-weight impurities at significant fraction of the main ion species, and Tungsten as heavy impurity in trace quantities. Next, in Sec.\,\ref{sec:21n}, we consider the special case of the $21N$-moment closure and compare it against the trace approximation, establishing a range of validity of the trace approximation for the four chosen cases. 
In Sec.\,\ref{sec:relaxation}, we seek to illustrate a few extensions to the closure by relaxing some of the assumptions from Sec.\,\ref{sec:closure}, such as for the linearized Burnett approximation, on the basis of which we build the closure for the drift approximation. 
The article is finally summarized in Sec.\,\ref{sec:summary}. We provide extensive appendices containing the details of the calculations performed, and in particular, provide explicit balance equations in Appendix \ref{sec:balanceequations}, first for the $13N$-moments (Appendix \ref{subsec:13n}), and then balance equations for general vectorial and tensorial moments (Appendix \ref{sec:generalbalanceequations}).   

\section{Moment-averaged kinetic equation}
\label{sec:lhs}

The Boltzmann equation which describes the evolution of the distribution function of a given species $\alpha$, $f_\alpha$ in the frame of the peculiar velocity $\mathbf{c_\alpha}=\mathbf{v_\alpha}-\mathbf{u}$, is given by
\begin{equation}
 \frac{d f_\alpha}{d t}+\mathbf{c_\alpha}.\nabla{f_\alpha}+\frac{1}{m_\alpha}\mathbf{F}^*_\alpha.\nabla_{c_\alpha}{f_\alpha} -c_{\alpha s}\frac{\partial f_\alpha}{\partial c_{\alpha_r}}\frac{\partial u_r}{\partial x_s}= \sum_{\beta} J_{\alpha\beta}, \label{boltzmannc}
\end{equation} 
where the common plasma flow velocity $\mathbf{u}$ is given by
\begin{equation}
 \rho \mathbf{u} = \sum_\alpha \rho_\alpha \mathbf{u_\alpha},\ \rho = \sum_\alpha \rho_\alpha,
\end{equation}
where $\rho$ represents the mass density.The $d/dt$ represents full time derivative given by $d/dt=\partial/\partial t+\mathbf{u}.\nabla$, and where the force term $\mathbf{F_\alpha}$ and $d\mathbf{u}/dt$ are combined to write the force in the moving frame $\mathbf{F^*_\alpha}=\mathbf{F_\alpha}-m_\alpha d\mathbf{u}/dt$. The LHS is referred to as the ``free-streaming term'', and the right hand side (RHS) $J_{\i\j}$ is the binary collision term $\i$ and every other species of the system. 
For any quantity $\psi_\alpha$ depending purely on species peculiar velocity $\mathbf{c}_\alpha$, one can average over Eq.~\ref{boltzmannc} which attains the following form
\begin{multline}
\frac{d}{dt}n_\alpha \langle\psi_\alpha\rangle+n_\alpha\langle\psi_\alpha\rangle\nabla.\mathbf{u} +\nabla.(n_\alpha\langle\psi_\alpha\mathbf{c_\alpha}\rangle)\\
-n_\alpha\left\{\left\langle\frac{d\psi_\alpha}{dt}\right\rangle +\langle \mathbf{c_\alpha}.\nabla\psi_\alpha\rangle +\frac{1}{m_\alpha}\langle\mathbf{F}^*_\alpha.\nabla_{c_\alpha}{\psi_\alpha}\rangle\right.\\
\left. -\left(\left\langle c_{\alpha s}\frac{\partial \psi_\alpha}{\partial c_{\alpha_r}}\right\rangle\frac{\partial u_r}{\partial x_s}\right) \right\} = R_\alpha, \label{eq:transport}
\end{multline}
where the $\psi_\i$-averaged RHS represents the moment of the collision operator.

At this juncture, following previous work, for a multi-temperature case, we choose Sonine-Hermite polynomials $G_\alpha^{mn}$ of the form,
\begin{multline}
 G_\alpha^{mn}(\mathbf{c_\alpha},\gamma_\alpha) = (-1)^n n! m_\alpha\gamma_\alpha^{-(n+m/2)}\\
 \times S^n_{m+1/2}\left(\frac{\gamma_\alpha}{2}\mathbf{c}^2_\alpha\right)P^{(m)}(\gamma_\alpha^{1/2}\mathbf{c_\alpha}),
 \label{eq:sonine-hermite}
\end{multline}
where $\gamma_\alpha={m_\alpha}/{kT_\alpha}$, with the Sonine polynomial $S^n_{m+1/2}$ given by
\begin{equation}
 S^n_{m+1/2}\left(\frac{\gamma_\alpha}{2}\mathbf{c}^2_\alpha\right) = \sum_{p=0}^n \left(-\frac{\gamma_\alpha}{2}\mathbf{c}^2_\alpha\right)^p\frac{(m+n+1/2)!}{p!(n-p)!(m+p+1/2)!},\label{eq:soninepolynomial}
\end{equation}
and the irreducible tensorial monomial $p^{(m)}$ given by
\begin{equation}
 P^{(m)}(\gamma_\alpha^{1/2}\mathbf{c_\alpha})  = \sum_{l=0}^{[m/2]} (-2)^l\gamma_\alpha^{m/2}\frac{(2m-2l)!m!}{(2m)!(m-l)!}\mathbf{c}^{m-2l}_\alpha\bm{\delta}^{l} c_\alpha^{2l},\label{eq:irreduciblemonomial}
\end{equation}
where the product $\mathbf{c}^{m-2l}_\alpha\bm{\delta}^{l}$ is symmetrized (see Appendix \ref{sec:symmetrizationnote} for a detailed explanation), such that $P^{(m)}$ is symmetric and traceless for $m>1$. 
The moments $b^{mn}_\alpha$ are calculated as
\begin{equation}
 n_\alpha b^{mn}_\alpha = \int G_\alpha^{mn} f_\alpha d\mathbf{c}_\alpha. 
 \label{eq:moment}
\end{equation}  
 
We now substitute $\psi_\i=G^{mn}_\i$ in Eq.\,(\ref{eq:transport}), from which we get the the moment-averaged kinetic equation in the form
\begin{multline}
\frac{d}{dt}n_\alpha \langle{G^{mn}_\i}\rangle+n_\alpha\langle{G^{mn}_\i}\rangle\nabla.\mathbf{u} +\nabla.(n_\alpha\langle\mathbf{c_\alpha}{G^{mn}_\i}\rangle)\\
-n_\alpha\left\{\left\langle\frac{d{G^{mn}_\i}}{dt}\right\rangle +\langle \mathbf{c_\alpha}.\nabla{G^{mn}_\i}\rangle\right.\\
\left.+\frac{1}{m_\alpha}\langle\mathbf{F}^*_\alpha.\nabla_{c_\alpha}{{G^{mn}_\i}}\rangle -\left(\left\langle c_{\alpha s}\frac{\partial {G^{mn}_\i}}{\partial c_{\alpha_r}}\right\rangle\frac{\partial u_r}{\partial x_s}\right) \right\}=\sum_\j R_{\i\j}^{mn}
\label{eq:transportgmn}.
\end{multline}
The collisional RHS $R_{\i\j}^{mn}$ is determined from a linearized collision operator using the following ansatz for the distribution function
\begin{equation}
f_\alpha = f_\alpha^{(0)}\sum_{m}^{m_{max}}\sum_{n}^{l_m} 2^{2n}m_\alpha^{-2}\gamma_\alpha^{2n+m} \tau_{mn} b^{mn}_{\alpha}G^{mn}_{\alpha},\label{eq:ansatz}
\end{equation}
where the distribution function at thermodynamic equilibrium $f_\alpha^{(0)}$ is chosen at the species specific temperature $T_\i$, given by
\begin{equation}
 f_\alpha^{(0)} =n_\alpha\left( \frac{\gamma_\alpha}{2\pi} \right)^{3/2}\exp{\left( -\frac{\gamma_\alpha}{2}c_\alpha^2\right)},
\end{equation}
and where the constant term $\tau_{mn}$ is given by
\begin{equation}
 \tau_{mn}= \frac{(2m+1)!(m+n)!}{n!(m!)^2(2m+2n+1)!}. \nonumber
\end{equation}
The number of moments $n$ can be considered to depend on rank-$m$ such that $n\equiv n(m)$. Thus, the number of moments desired can be chosen for each rank-$m$ by fixing the value of $l_m$. E.g. $m_{max}=1,l_0=1,l_1=0$ will lead to the $5N$-moment system of equations, $m_{max}=2,l_0=1,l_1=1,l_2=0$ will lead to the $13N$-moment system, and $m_{max}=2,l_0=1,l_1=2,l_2=1$ leads to the $21N$-moment system of equations and so on. We shall use the notation $n(m)$ to clarify the rank dependence of the number of moments wherever necessary. In our case following Zhdanov et al, we used the linearized Boltzmann collision operator as follows
\begin{multline}
 R_{\alpha\beta}^{mn}\approx\iiint f^{(0)}_\alpha f^{(0)}_\beta(G^{mn\prime}_\i-G^{mn}_\i)(1+\Phi_\alpha+\Phi_\beta)\\
 \times  g\sigma_{\alpha\beta}(g,\chi)d\Omega d\mathbf{c_{\alpha}}d\mathbf{c_{1\beta}}, \label{eq:momentaveragefirstfrom}
\end{multline}
where for species $\i$, $\Phi_\alpha$ is given by
\begin{equation}
 \Phi_\alpha=\sum_{(m,n(m))\neq (0,0)} 2^{2n}m_\alpha^{-2}\gamma_\alpha^{2n+m} \tau_{mn} b^{mn}_{\alpha}G^{mn}_{\alpha}
\end{equation}
This leads to a moment-averaged collision term of the form
\begin{equation}
 R_{\alpha\beta}^{mn}=\sum_l R_{\alpha\beta}^{mnl},
\end{equation}
where 
\begin{equation}
 R_{\alpha\beta}^{mnl}=(1-\delta_{m0}\delta_{l0})(A_{\alpha\beta}^{mnl}b^{ml}_\alpha+B_{\alpha\beta}^{mnl}b^{ml}_\beta)+\delta_{m0}\delta_{l0}C_{\alpha\beta}^{mnl}
 \label{eq:collision_second_form}.
\end{equation}
The coefficients $A_{\alpha\beta}^{mnl}$, $B_{\alpha\beta}^{mnl}$ and $C_{\alpha\beta}^{mnl}$ are dependent only on the masses, temperatures, number densities, and the potential of interaction between the two species $\i$ and $\j$. The expressions for them, alongwith the method of derivation, can be found in the previous article by the authors. 
In addition, Zhdanov et al provide a set of single-temperature coefficients at the plasma common temperature in Ref.\,\onlinecite{yushmanov_diffusion_1980,zhdanov_pfirschschlueter_1977,zhdanov_transport_2002} for the $21N$-moment scheme, which in our previous article, we were able to generalize the calculation for any arbitrary number of moments (we also choose the value of the factor $d_{\i\j}=\gamma_{\i\j}/2$. See Ref.\,\cite{raghunathan_generalized_2021} and Appendix \ref{sec:dij} for details). For both the multi-temperature and single temperature collision coefficients, we choose the cross section expressions for the shielded Coulomb potential. 
Note however that the multi-temperature coefficients are generally valid for any temperature differences, but because of the form of the ansatz Eq.\,(\ref{eq:ansatz}), there in an assumption that the flow velocities of all species are close to plasma common flow velocity, i.e.
\begin{equation}
 |\mathbf{u}_\i-\mathbf{u}|\ll \left(\frac{kT_\i}{m_\i}\right)^{1/2}.
\end{equation}. 

Returning to the LHS, the force is decomposed into general non-velocity dependent body force $\mathbf{X}$, and the electromagnetic force as follows
\begin{equation}
 \mathbf{F}^*_\alpha=\mathbf{X}_\i+Z_\i e(\mathbf{E}+\mathbf{v}_\i\times\mathbf{B})-m\frac{d\mathbf{u}}{dt},
\end{equation}
and therefore, the force dependent term can be written as
\begin{multline}
 \frac{1}{m_\alpha}\langle\mathbf{F}^*_\alpha.\nabla_{c_\alpha}{{G^{mn}_\i}}\rangle=\left(\frac{X_{\i l}}{m_\i}+\frac{Z_\i eE_{\i l}}{m_\i}\right.\\
 \left.+\frac{Z_\i e}{m_\i}\{\mathbf{u}\times\mathbf{B}\}_l-\frac{d u_l}{dt}\right)\left\langle\frac{\partial G^{mn}_\i}{\partial c_{\i l}} \right\rangle,
\end{multline}
where the term $\mathbf{E}+\mathbf{u}\times\mathbf{B}$ can be thought of as the usual electric field in the moving frame $\mathbf{E^*}$. The magnetic field term can be written as
\begin{equation}
 \frac{Z_\i e}{m_\alpha}\langle\mathbf{c}_\alpha\times\mathbf{B}.\nabla_{c_\alpha}{{G^{mn}_\i}}\rangle=\epsilon_{rst}\omega_{\i t}\left\langle c_{\i s}\frac{\partial G^{mn}_\i}{\partial c_{\i r}}   \right\rangle
\end{equation}
where $\epsilon_{rst}$ is the Levi-Civita tensor and $\bm{\omega}_\i=Z_\i e\mathbf{B}/m_\i$. One can also expand the average of the time derivative term as follows
\begin{equation}
 \left\langle\frac{d{G^{mn}_\i}}{dt}\right\rangle=-\frac{\gamma_\i}{T_\i}\left\langle\frac{\partial {G^{mn}_\i}}{\partial\gamma_\i}\right\rangle\frac{\partial T_\i}{\partial t}-\frac{\gamma_\i}{T_\i}\left\langle\mathbf{c}_\i\frac{\partial  {G^{mn}_\i}}{\partial\gamma_\i}\right\rangle\cdot\nabla T_\i.
\end{equation}
It is clear from these expressions that one needs to obtain the various $G^{mn}$ derivatives, contractions, etc beforehand in order to fully expand the LHS. One can use the fact that the irreducible representation of Hermite polynomials is a product of the Sonine polynomial and the irreducible tensorial monomial, and use their properties, such as recurrence, contraction with a vector and rank-2 tensor (1-fold and 2-fold inner products), single and double derivative, to derive the properties for $G^{mn}$. These properties are reproduced in the appendix \ref{sec:polynomialidentities} (with some changes for consistency) for the ease of reference.

On using these values and substituting them in Eq.\,(\ref{eq:transportgmn}), the following expression for the balance equation for the general $(mn)^{th}$ moment is found
\begin{widetext}
 \begin{multline}
 \frac{d}{dt}(n_\alpha {b^{mn}_\i})+n_\alpha{b^{mn}_\i}\nabla\cdot\mathbf{u}
 +\frac{\partial}{\partial x_r}\left[n_\i b^{m+1,n}_{\i r} +\frac{n}{\gamma_\i}n_\i b^{m+1,n-1}_{\i r}+\frac{2}{2m+1}\left(\frac{2m+2n+1}{2\gamma_\i}\{n_\i b^{m-1,n}_\i\bm{\delta}\}_r +\{n_\i b^{m-1,n+1}_\i\bm{\delta}\}_r     \right) \right]\\
 -\left(\frac{X_{\i r}}{m_\i}+\frac{Z_\i eE_{r}}{m_\i}+\{\mathbf{u}\times\bm{\omega}_\i\}_r-\frac{d u_r}{dt}\right)\left[nn_\i b^{m+1,n-1}_{\i r} +\frac{2m+2n+1}{2m+1}\{n_\i b^{m-1,n}_\i\bm{\delta}\}_r \right]\\
 +n_\i\left(\frac{\partial u_r}{\partial x_s}-\epsilon_{rst}\omega_{\i t} \right)\left[ nb^{m+2,n-1}_{\i rs} +\frac{n(n-1)}{\gamma_\i}b^{m+2,n-2}_{\i rs}+\frac{2n}{2m+3}\bm{\delta}b^{mn}_\i+\frac{n(2m+2n+1)}{(2m+3)\gamma_\i}\bm{\delta}b^{m,n-1}_\i+\{\bm{\delta} b^{mn}_{\i s}\}_r\right.\\
   \left.-\frac{2n}{2m-1}\bm{\delta}b^{mn}_{\i rs}-\frac{n(2m+2n+1)}{(2m-1)\gamma_\i}\bm{\delta}b^{m,n-1}_{\i rs}+\frac{2m+2n+1}{4m^2-1}\left(2\{b^{m-2,n+1}_\i\bm{\delta\delta}\}_{rs}+\frac{2m+2n-1}{\gamma_\i}\{b^{m-2,n}_\i\bm{\delta\delta}\}_{rs} \right) \right]\\
   +\frac{n_\i k}{2m_\i}n(2m+2n+1)b^{m,n-1}_\i\frac{d T_\i }{dt}\\
   +\frac{n_\i k}{2m_\i}n(2m+2n+1)\left[b^{m+1,n-1}_{\i r}+\frac{n-1}{\gamma_\i}b^{m+1,n-2}_{\i r}+\frac{2}{2m+1}\left(\frac{2m+2n-1}{2\gamma_\i}\{b^{m-1,n-1}_\i\bm{\delta}\}_r + \{b^{m-1,n}_\i\bm{\delta}\}_r\right) \right]\frac{\partial T_\i}{\partial x_r}\\
   =R^{mn}_\i=\sum_\j \sum_l [(1-\delta_{m0}\delta_{l0})(A_{\alpha\beta}^{mnl}b^{ml}_\alpha+B_{\alpha\beta}^{mnl}b^{ml}_\beta)+\delta_{m0}\delta_{l0}C_{\alpha\beta}^{mnl}]. \label{eq:generallhs}
\end{multline}
\end{widetext}
This equation represents the most general fluid moment obtained for a moment average with $G^{mn}_\i$. The first line of the equation contains the convective time derivatives $d/dt$ and the space derivatives of the moments (through gradients of scalars and divergences of tensors). The second line contains the velocity-independent force dependent terms. The third and fourth lines contain the viscous-stress and magnetic field dependent terms. The fifth and sixth lines contain the temperature time gradient $dT_\i/dt$ and space gradient $\nabla T_\i$ dependent terms. The RHS contains the moment averaged collision operator. The general expressions for the RHS $R_{\i\j}^{mn}$ for the linearized moment-averaged Boltzmann collision operator for a multi-temperature case, for up to rank-2 moments, can be found in the previous work by the authors\cite{raghunathan_generalized_2021}, which takes the form of a linear combination of moments of a similar rank-$m$. 

In this expression, all terms are symmetrized (See Appendix \ref{sec:symmetrizationnote}), and repeated indices are summed over.
It has to be also mentioned that any $r^{th}$ or $(rs)^{th}$ component of a tensor mentioned refers to the additional ranks the tensor has, i.e. the rank-0 quantities do not possess an  $r^{th}$ component and rank-1 quantities do not possess an additional $(rs)^{th}$ component. Should such a case arise, the term may be safely set to zero.

The above expression also differs from the general expression  Eq.\,(A1.7) given in Zhdanov et al in Ref.\,\onlinecite{zhdanov_transport_2002}, who have a much simpler magnetic field term of the form $n_\i b^{mn}_{\i s}\epsilon_{rst}\omega_{\i t} \bm{\delta}_r$. This is just a cosmetic difference, because the Levi-Civita tensor is antisymmetric, and on being contracted with any symmetric term, would vanish. The only term that survives on the expansion is $n_\i \epsilon_{rst}\omega_{\i t}\{ b^{mn}_{\i s}\bm{\delta}\}_r$, which can then be expanded out as $n_\i \bm{\delta}_rb^{mn}_{\i s}\epsilon_{rst}\omega_{\i t}$. 
However, we also also have an additional term $\mathbf{u}\times\bm{\omega}$ term (which could alternately be written as a electric field in the moving frame). Note also the difference in the coefficient of $\bm{\delta}b_{\i rs}^{mn}$, and the presence of the Boltzmann constant $k$ multiplying the coefficients of $dT_\i/dt$ and $\nabla T_\i$. 
This expression is similar to the one found in Eq.\,(3.1.3) of Ref.\,\onlinecite{weinert_multi-temperature_1982}, but defined at the species specific temperature using the Sonine-Hermite polynomials instead of the spherical harmonics.

\section{Multi-temperature generalized Zhdanov closure}
\label{sec:closure}

As one can notice from Eq.\,(\ref{eq:generallhs}), the moment equation for a general moment $b^{mn}$ of rank-$m$ , contains quantities which are of ranks $m\pm 1$ and $m\pm2$. This implies that if one truncates the series by choosing a certain number of moments $M$ in the distribution function, one will obtain $N$ corresponding balance equations, but the number of variables in these balance equations will exceed $M$. This leaves the set of $M$ equations unclosed, requiring elimination of the excess variables through some means. Furthermore, generally most SOL/edge fluid packages solve only for the plasmadynamical quantities, i.e. density, temperature/pressure/energy, and flow momentum density, i.e. $(\rho,p/T/E,\rho\mathbf{w})$, which constitute $5N$-moments, where $N$ is the number of species. Generally, all additional variables other than these three must be eliminated.

One such direct method is the closure recommended by Grad himself, which involves calculating the higher order moments in terms of lower ones by using the ansatz for the distribution function Eq.\,(\ref{eq:ansatz}) in the expression for the moment Eq.\,(\ref{eq:moment}). For symmetric and irreducible moments, assuming the ansatz of the distribution function does not contain the moment, such a process is equivalent to setting to zero that moment. It has come to be known as Grad's closure in literature, and is the finishing touch in the description of Grad's method. This generally works well for fluids such as monoatomic gases, where the RHS vanishes in the hydrodynamic equations and only the LHS needs to be treated. In particular, in a $5N$-moment approximation, the heat-flux moment $b^{10}_\i$, i.e.\,$\mathbf{h_\i}=0$ (thus the conventional heat flux $\mathbf{q}_\i=(5/2)p_\i \mathbf{w}_\i$) and the stress tensor $\pi_\i=0$ from Grad's closure. One can notice that it does not recover the temperature-gradient force term in RHS of the momentum balance equation, because heat-flux $b^{11}_\i$ calculated in this manner does not depend on the temperature gradients. It also does not recover the usual visco-elastic form of the stress tensor, where the stress tensor is proportional to the rate-of-strain tensor.  Thus, this leaves us with an oversimplified description of the heat-flux and the stress tensor if the $13N$-moment Grad's closure is used. Also notice that this closure typically leads to the higher-moment of any species depending only on the lower moments of that species only.

One other method was proposed by a series of authors\cite{landshoff_convergence_1951,spitzer_transport_1953,kaneko_transport_1960,braginskii_transport_1965,kaneko_electrical_1978,kaneko_thermal_1980, balescu_transport_1988}
, under what is known as the linear transport assumptions, to find approximate equations for the higher-order moments, usually using the first Chapman-Enskog approximation, and ignoring some additional terms on the basis of low electron mass, for a single-temperature plasma, to find transport coefficients for an ion-electron plasma. These usually involve a simultaneous solution of the approximate linearized equations for ions and electrons, which after the approximations may or may not be coupled, to find the transport coefficients. Some of these works use the linearized Landau collision operator, and the others the linearized Boltzmann operator. These works furthermore find the usual form of the friction and thermal forces, in terms of the flows and the temperature gradients respectively. However, the transport coefficients and the forces calculated from such a scheme, which is only strictly applicable to an ion-electron single-temperature plasma, are not usually applicable to a plasma with impurities at different charge states in significant amounts. Thus, to address this problem, Zhdanov et al proposed a new solution scheme for the case of impurities in a plasma.

\subsection{General assumptions}

Zhdanov et al proposed a new scheme\cite{zhdanov_pfirschschlueter_1977,yushmanov_diffusion_1980} that involves linearizing the balance equation for the higher-order moments under certain assumptions as follows
\begin{enumerate}
\item Firstly, a desired set of moments is chosen, comprising of plasma dynamical moments $(\rho,\rho\mathbf{w},nkT)$, thermodynamically privileged higher-order moments $(\mathbf{h},\pi)$, and thermodynamically non-privileged moments of even higher-order. The balance equations for this set of moments is calculated with the collision terms being calculated by the ansatz of the distribution function containing these moments Eq.\,(\ref{eq:ansatz}).
 \item For moments out of this desired set, Grad's closure is used on them, i.e. for our case of symmetric irreducible moments, these higher-order moments are set to zero.
 \item The plasma dynamical moment $(\rho,\rho\mathbf{w},nkT)$, are considered zeroth order in $K_n$. Their space gradients are considered again to go as first order in $K_n$, where
 \begin{equation}
  K_n\sim \frac{\lambda}{L_{sc}},\  \frac{\tau}{\tau_{sc}},
 \end{equation}
 where $\lambda$ represents mean free path between collisions, and $\tau$ represents mean time between collisions, and $L_{sc}$ and $\tau_{sc}$ represents the scale lengths and the scale times of the system in question. However $(\mathbf{w},T)$ are higher than first order, the products of $(\mathbf{w},T)$ with themselves or other moments may be considered to be of order higher than one in $K_n$.
 \item The higher-order moments, both privileged and non-privileged, are considered to be of the order of one or higher in Knudsen number $K_n$.
 \item The time derivatives of higher-order moments, both privileged and non-privileged, are neglected, meaning that quantity that the higher-order moment represents changes slowly over the characteristic timescale of the system $\tau_{sc}$. This essentially means that the time derivatives of higher-order moments are considered larger than first order in Knudsen number $K_n$. Physically, it means that the moment evolves slower than any changes in the moment caused by collisions.
 \item The space gradients of non-privileged higher-order moments are also neglected, which means that the macroscopic quantities represented by the higher-order moments change gently over the scale length of the system $L_{sc}$, meaning the gradients of the higher-order quantities are of an order larger than one in $K_n$.
 \item Products of a higher-order moment with other moments, higher or lower order, are also neglected, since the product of moments which are the first order in Knudsen number will lead to quantities which are second order or higher in $K_n$.
 \item  The common flow $\mathbf{u}$ is considered zeroth order in $K_n$, but its space and time gradients are considered between order zero and one in $K_n$.
 \item Consequently, $\gamma$ can be considered to go as order one, and factors of $\gamma^{-1}$ may be considered to be between $(-1)^{st}$ in $K_n$. Factor of $k/m$ may be considered even lower than the $(-1)^{st}$ order. (Which would make some gradients like $\gamma^{-1}\nabla\mathbf{w}$ and $(k/m)\nabla T$ equal to or lower than order one.)
\end{enumerate}
These approximations essentially transform the balance equations of these higher-order moments into linear, non-differential equations, which can then be solved to obtain approximate values of these higher-order moments. It is similar to the first-order approximation of the Chapman-Enskog scheme, because of the explicit ordering in terms of the Knudsen number. (Note however that it is not exactly the same as the Burnett approximation. We shall touch lightly on this point later).  A similar scheme has also been proposed by Balescu\cite{balescu_transport_1988}, which retains the time derivatives of the higher-order moments however, in addition to the explicit ordering in terms of a provisionally defined hydrodynamic timescale $\tau_H$ rather than  the Knudsen number.  

\subsection{Specific assumptions in the reduction of the balance equations}

On applying the approximations as outlined, we first find that the balance equations for the plasmadynamical quantities $(\rho,\rho\mathbf{w},nkT)$ survive as it is, as can be seen from Appendix \ref{subsec:13n}. However, in the LHS of these balance equations, we find that there are open variables such as the heat-flux $\mathbf{h}_\i$ and the stress tensor $\pi_\i$. Again, on applying the approximations from the previous subsection, we find the balance equations for the heat-flux $\mathbf{h}_\i$ and the stress tensor $\pi_\i$ reduce to the following
 \begin{multline}
 -\pi_{\i l}\left(\frac{X_{\i l}}{m_\i}+\frac{Z_\i eE_{\i l}}{m_\i}+\{\mathbf{u}\times\bm{\omega}_\i\}_l-\frac{d u_l}{dt}\right)\\
   -\frac{5}{2}\frac{kT_\i}{m_\i}\nabla.{\pi_{\i}}+\frac{5}{2} \frac{k}{m_\i}n_\i kT_\i \frac{\partial T_\i}{\partial x_r}\\
   -\mathbf{h}_\i\times\bm{\omega}_\i=\sum_\j R^{11}_{\i\j}, \label{eq:13nhlinear}
\end{multline}
and
 \begin{multline}
 \frac{4}{5}\left(\frac{5}{2}n_\i kT_\i\left\{ \frac{\partial {w}_{\i r}}{\partial x_s}\right\}+\left\{ \frac{\partial {h}_{\i r}}{\partial x_s}\right\}\right) \\
 -\rho_\i\left\{{w}_{\i r}\left(\frac{X_{\i s}}{m_\i}+\frac{Z_\i eE_{\i s}}{m_\i}+\{\mathbf{u}\times\bm{\omega}_\i\}_s-\frac{d u_s}{dt}\right)\right\} \\
 +2n_\i kT_\i\left\{\frac{\partial u_r}{\partial x_s}\right\}-2\{\pi_{\i sl}\epsilon_{rst}\omega_{\i t}\}=\sum_\j R_{\i\j}^{20}\label{eq:13npilinear}
\end{multline}
respectively. Furthermore, from the RHS of the balance equations, we can observe that there are higher-order moments which remain unclosed. Therefore, the balance equations for the general higher-order vectorial and tensorial non-privileged moments $\mathbf{b}^{1n(1)}_\i$ and $b^{2n(2)}_\i$  can be similarly reduced to
\begin{multline}
-n(1)n_\i b^{2,n(1)-1}_{\i l}\left(\frac{X_{\i l}}{m_\i}+\frac{Z_\i eE_{\i l}}{m_\i}+\{\mathbf{u}\times\bm{\omega}_\i\}_l-\frac{d u_l}{dt}\right)\\
  -\frac{2n(1)+3}{3}n_\i b^{0,n(1)}_\i\left(\frac{\mathbf{X}_{\i}}{m_\i}+\frac{Z_\i e\mathbf{E}}{m_\i}+\mathbf{u}\times\bm{\omega}_\i-\frac{d \mathbf{u}}{dt}\right) \\
 -\mathbf{b}^{1n(1)}_\i\times\bm{\omega}_\i=\sum_\j R^{1n(1)}_{\i\j},
\end{multline}
and
\begin{multline}
 -\frac{2n(2)+5}{5}n_\i b^{1,n(2)}_\i \left(\frac{\mathbf{X}_{\i }}{m_\i}+\frac{Z_\i e\mathbf{E}}{m_\i}+\mathbf{u}\times\bm{\omega}_\i-\frac{d \mathbf{u}}{dt}\right)\\
  -2\{b^{2n(2)}_{\i sl}\epsilon_{rst}\omega_{\i t}\}=\sum_\j R_{\i\j}^{2n(2)}
\end{multline}

In the absence of body forces $\mathbf{X}_\i$ and parallel electric fields $E_\parallel$, and a static common flow, these equations for the heat flux $\mathbf{h}_\i$ and stress tensor $\pi_\i$ can be further resolved along the direction parallel to the magnetic field, such that we obtain
 \begin{equation}
  -\frac{5}{2}\frac{kT_\i}{m_\i}\nabla_{\parallel}.{\pi_{\i}}+\frac{5}{2} \frac{k}{m_\i}n_\i kT_\i \nabla_\parallel T_\i=\sum_\j R^{11}_{\i\j\parallel}, \label{eq:13nhlinearnofield}
\end{equation}
and
 \begin{multline}
  \frac{4}{5}\left(\frac{5}{2}n_\i kT_\i\left\{ \frac{\partial {w}_{\i r}}{\partial x_s}\right\}_{\parallel\parallel}+\left\{ \frac{\partial {h}_{\i r}}{\partial x_s}\right\}_{\parallel\parallel}\right) \\
 2n_\i kT_\i\left\{\frac{\partial u_r}{\partial x_s}\right\}_{\parallel\parallel}=\sum_\j R_{\i\j\parallel\parallel}^{20}\label{eq:13npilinearnofield}
\end{multline}
respectively, where for a vector $\mathbf{a}$, $a_\parallel=\mathbf{{b}}\mathbf{{b}}.\mathbf{a}$ and for a traceless symmetric tensor $A$, $A_{\parallel\parallel}=(\mathbf{{b}}\mathbf{{b}}-\bm{\delta}/3)(\mathbf{{b}}\mathbf{{b}}-\bm{\delta}/3):A$, where $\mathbf{b}$ is a unit vector along the direction of the magnetic field. 
Now, one can further assume that the parallel gradients of the higher-order moments are weak in nature, and reduce these equations further down to
 \begin{equation}
 \frac{5}{2} \frac{k}{m_\i}n_\i kT_\i \nabla_\parallel T_\i=\sum_\j R^{11}_{\i\j\parallel}, \label{eq:13nhlinearnofield1}
\end{equation}
and
 \begin{equation}
  2n_\i kT_\i\left\{\frac{\partial u_r}{\partial x_s}\right\}_{\parallel\parallel}=\sum_\j R_{\i\j\parallel\parallel}^{20}\label{eq:13npilinearnofield1}.
\end{equation}
(Such an assumption may not be valid under certain conditions, to which we will return to in Sec. \ref{sec:relaxation}). 
Under the same assumptions of no body forces and no parallel electric fields, the balance equations for the non-privileged higher-order moments in the parallel direction can be reduced as follows
$\mathbf{b}^{1n(1)}_\i$ and $b^{2n(2)}_\i$  (see Appendix \ref{sec:generalbalanceequations}) can be similarly reduced to 
\begin{equation}
 0=\sum_\j R^{1n(1)}_{\i\j\parallel},
\end{equation}
and
\begin{equation}
 0=\sum_\j R_{\i\j\parallel\parallel}^{2n(2)}\label{eq:generalmomenteqrank2}.
\end{equation}
The expressions Eqs.\,(\ref{eq:13nhlinearnofield1})-(\ref{eq:generalmomenteqrank2}) represent then a closed set of equations which form the basis of the generalized Zhdanov closure.

\subsection{Solution of the linear reduced system of balance equations}

We now proceed to explain how to solve the linear system of equations given by Eqs.\,(\ref{eq:13nhlinearnofield1})-(\ref{eq:generalmomenteqrank2}). First, we define ${M^{mnl}_{\alpha\gamma}}$ such that
 \begin{equation}
 M^{mnl}_{\alpha\gamma}=\left\{\begin{tabular}{ll}
             $A_{\i\i}^{mnl}+B_{\i\i}^{mnl}+\sum_{\j\neq\i}A_{\i\j}^{mnl}$ & $,\ \i=\gamma$\\
             $B_{\i\gamma}^{mnl}$ & $,\ \i\neq\gamma$
             \end{tabular}\right. 
             \label{eq:mmnl}
\end{equation}
where the index $\gamma$ runs over all $N$ species. It is also similar to the notations $q^{nl},p^{nl}$ used in Refs.\,\onlinecite{devoto_simplified_1967,devoto_thesis,bonnefoi_thesis_1975,bonnefoi_thesis_1983,rat_transport_2001}. 
We then define an $N\times N$ matrix $M^{mnl}$ 
\begin{equation}
 M^{mnl}=\left(\begin{tabular}{cccc}
          $M^{mnl}_{\alpha\gamma}$ & $M^{mnl}_{\alpha\delta}$ & $\ldots$  & $M^{mnl}_{\alpha\omega}$\\
          $M^{mnl}_{\delta\alpha}$ & $\ddots$ & $ $ & $\vdots$\\
          $\vdots$ & $ $ & $\ddots$ & $\vdots$\\
          $M^{mnl}_{\omega\alpha}$ & $\ldots$ & $\ldots$ & $M^{mnl}_{\omega\omega}$
         \end{tabular}\right), \label{eq:bstar200}
\end{equation}
such that
\begin{equation}
 R^{mn}_{\i}=\sum_l M^{mnl}_\i B^{ml}_\i, 
\end{equation}
where $M^{mnl}_\i$ is the row-$\i$, and where $B^{mnl}$ is a column vector of length $N$ given by 
 \begin{equation}
 B^{ml}=\left(\begin{tabular}{c}
          $b^{ml}_\i$ \\
          $b^{ml}_\gamma$\\
          $\vdots$ \\
          $b^{ml}_\omega$
         \end{tabular}\right). \label{eq:BMN}
\end{equation}
Note that, for the special case of $m=0,\ l=1$, 
\begin{equation}
 R^{0n}_{\i}=\sum_\beta C^{0n0}_{\i\beta}+\sum_{l=1} M^{0nl}_\i B^{0l}_\i. 
\end{equation}

For the rank-1 quantities, we define an $(l_1-1)\times(l_1-1)$ block matrix such that,   
\begin{equation}
 M^{1}_{l_1}=\left(\begin{tabular}{cccc}
          $M^{111}$ & $M^{112}$ & $\ldots$  & $M^{11l_1}$\\
          $M^{121}$ & $\ddots$ & $ $ & $\vdots$\\
          $\vdots$ & $ $ & $\ddots$ & $\vdots$\\
          $M^{1l_11}$ & $\ldots$ & $\ldots$ & $M^{1l_1 l_1}$
         \end{tabular}\right),
\end{equation}
and a block column vector given by 
\begin{equation}
 M^{1}_{0l_1}=\left(\begin{tabular}{c}
          $M^{110}$ \\
          $M^{120}$\\
          $\vdots$ \\
          $M^{1l_1 0}$
           \end{tabular}\right),
\end{equation}
such that the linear transport relations can be represented as
\begin{equation}
 \Lambda^{1}_{l_1} T_{\parallel}=M^{1}_{0l_1}W_\parallel+M^{1}_{l_1}H^{1}_{l_1 \parallel},\label{eq:closuregeneralrank1}
\end{equation}
where  $H^{1}_{l_1 \parallel}$ and $ \Lambda^{1}_{l_1}$ are block column vectors of length $(l_1-1)$ given by
 \begin{equation}
 H^{1}_{l_1 \parallel}=\left(\begin{tabular}{c}
          $H_\parallel$ \\
          $B^{12}_\parallel$\\
          $\vdots$ \\
          $B^{1l_1}_{\parallel}$
         \end{tabular}\right),\
     \Lambda^{1}_{l_1}=\left(\begin{tabular}{c}
          $\Lambda$ \\
          $0$\\
          $\vdots$ \\
          $0$
         \end{tabular}\right),        
\end{equation}
where $T_{\parallel}$, $W_{\parallel}$, $H_\parallel$, $B^{1l}_\parallel$,  are column vectors of length $N$ containing values of $\nabla_\parallel T_\gamma$, $\rho_\gamma w_{\gamma\parallel}$, $h_{\gamma\parallel}$, $b^{1l}_{\gamma\parallel}$ respectively, and $\Lambda$ is a $N\times N$ diagonal matrix containing $\frac{5}{2} \frac{k}{m_\gamma}n_\gamma kT_\gamma$ on its diagonal.

Similarly, for rank-2 quantities, a $l_2\times l_2$ block matrix $M^{2}_{l_2}$ can be defined as follows
\begin{equation}
 M^{2}_{l_2}=\left(\begin{tabular}{cccc}
          $M^{200}$ & $M^{201}$ & $\ldots$  & $M^{20l_2}$\\
          $M^{210}$ & $\ddots$ & $ $ & $\vdots$\\
          $\vdots$ & $ $ & $\ddots$ & $\vdots$\\
          $M^{2l_20}$ & $\ldots$ & $\ldots$ & $M^{2l_2l_2}$
         \end{tabular}\right),
\end{equation}
from which the set of linear transport relations for rank-2 quantities can now be represented as 
\begin{equation}
 -2P^2_{l_2}\epsilon_{\parallel\parallel}=M^{2}_{l_2}\Pi^{2}_{l_2\parallel\parallel}\label{eq:closuregeneralrank2}
\end{equation}
where $\Pi^{2}_{l_2\parallel\parallel}$ and $P^{2}_{l_2}$ are block column vectors of length $l_2$
 \begin{equation}
 \Pi^{2}_{l_2\parallel\parallel}=\left(\begin{tabular}{c}
          $\Pi_{\parallel\parallel}$ \\
          $B^{2l}_{\parallel\parallel}$\\
          $\vdots$ \\
          $B^{2l_2}_{\parallel\parallel}$
         \end{tabular}\right),\
     P^{2}_{l_2}=\left(\begin{tabular}{c}
          $P$ \\
          $0$\\
          $\vdots$ \\
          $0$
         \end{tabular}\right),
\end{equation}
where  $\Pi_{\parallel\parallel}$, $P$ are column vectors of length $N$ containing values of $\pi_{\gamma\parallel\parallel}$ and $n_\gamma kT_\gamma$ respectively. The symbol $\epsilon_{\parallel\parallel}$ is shorthand for $\{\nabla\mathbf{u}\}$. 

The solutions to Eqs\,(\ref{eq:closuregeneralrank1}) and (\ref{eq:closuregeneralrank2}) are given by inverting the equations as follows
\begin{align}
H^{1}_{l_1 \parallel} &=(M^{1}_{l_1})^{-1}\Lambda^{1}_{l_1 } T_{\parallel}-(M^{1}_{l_1})^{-1}M^{1}_{0l_1}W_{\parallel}\label{eq:generalizedrank1}\\
 \Pi^{2}_{l_2\parallel\parallel}&=-2(M^{2}_{l_2})^{-1}P^{2}_{l_2}\epsilon_{\parallel\parallel}\label{eq:generalizedrank2}.
\end{align}
Thus one can obtain expressions for rank-1 general moments in terms of the flow velocities $w_\parallel$ and the temperature gradients $\nabla_\parallel T$, and expressions for rank-2 such that it is expressed in terms of the longitudinal rate-of-strain tensor $2\epsilon_{\parallel\parallel}$. One can confirm that the higher order moments calculated in this manner respect the orderings as mentioned in the previous subsection, for example, from Eq.\,(8.4.6) of Ref.\,\onlinecite{zhdanov_transport_2002}, where the heat flux term $\mathbf{h}_\i$ is a factor of $\gamma^{-1}_\i\tau_\i$ smaller than the flows and the temperature gradients, i.e. one order higher in $K_n$. The gradients of such terms will be one order higher than $K_n$ because of the assumptions on the gradients of the plasmadynamical quantities. Thus, the ordering chosen is generally respected by the results obtained from the closure.

Furthermore, the partial parallel thermal conductivities $\lambda_{\delta\gamma\parallel}$ can be found in the $(1,1)$ element of the block matrix $(M^{1}_{l_1})^{-1}\Lambda^{1}_{l_1 }$ and similarly, the partial longitudinal viscosities $\eta_{\gamma\parallel\parallel}$ can be found in the first element of the block column vector $(M^{2}_{l_2})^{-1}P^{2}_{l_2}$. The full parallel thermal conductivities $\lambda_{\delta\parallel}$ and full longitudinal viscosity $\eta_{\parallel\parallel}$ can be obtained by summing over the columns of the respective matrix elements, i.e.\,$\lambda_{\delta\parallel}=\sum_\gamma \lambda_{\delta\gamma\parallel}$, and $\eta_{\parallel\parallel}=\sum_\gamma \eta_{\gamma\parallel\parallel}$ respectively.

A version of this closure for $13N$-moment case was illustrated in the previous article by the authors, however, neglecting the self-collisions, for a purely non-magnetic case\cite{raghunathan_generalized_2021}. The non-magnetic case is superficially equivalent to the case here where we only consider parallel and longitudinal transport coefficients. 

\subsection{Friction and thermal forces}

Once the values of the higher-order rank-1 moments have been calculated, they can be substituted back into the RHS block of the momentum equation, i.e. the balance equation for $\rho_\i w_{\i \parallel}$, for all species as follows
\begin{equation}
 R^{00}_{\parallel}=M^{100}W_\parallel+M^{1}_{l_1 0}H^{1}_{l_1 \parallel}
\end{equation}
where $H^{1}_{l_1 \parallel}$ is given by Eq.\,(\ref{eq:generalizedrank1}), and a block row vector $M^{1}_{l_1 0}$ is given by
\begin{equation}
 M^{1}_{l_1 0}=\left(\begin{tabular}{cccc}
          $M^{101}$ & $M^{102}$ & $\ldots$  &$M^{10 l_1}$\\
         \end{tabular}\right).
\end{equation}
On substituting $H^{1}_{l_1 \parallel}$, we get
\begin{equation}
 R^{00}_{\parallel}=[M^{100}-M^{1}_{l_1 0}(M^{1}_{l_1})^{-1}M^{1}_{0l_1}]W_\parallel+M^{1}_{l_1 0}(M^{1}_{l_1})^{-1}\Lambda^{1}_{l_1 } T_{\parallel}.
\end{equation}
The part of the collisional force dependent on the flows $W_\parallel$ is termed the ``friction force'' and the part dependent on the temperature gradients $T_{\parallel}$ is termed the ``temperature-gradient force'' (or, ``thermal force'' at times)\cite{braginskii_transport_1965,stangeby_plasma_2000}.
As one can observe, the addition of terms from $H^{1}_{l_1 \parallel}$  ``augments'' the friction force term depending on $W_\parallel$, which has more contribution the higher the number of moments chosen.

\section{Convergence of the generalized Zhdanov closure}
\label{sec:tests}

In order to test the closure scheme, following the previous article, we investigate four cases of three-component plasmas, i.e. electrons and two other species, the most relevant to SOL/edge physics, as follows
\begin{itemize}
 \item The fusion fuel, containing deuterium and tritium (D-T), at comparable densities, as D-T fusion is planned to be used in current and future burning plasma campaigns, 
 \item Light impurities at significant fraction (10\%) of the main fuel species, i.e.\, hydrogen and carbon (C-H), with the carbon in the plasma originating from facing plasma components made of graphite, 
 \item Injected mid-weight impurities with densities at a small fraction (1\%) of the fuel species density, e.g.\,hydrogen and argon (Ar-H), often used for controlled experimentation with impurities, for measurement of background plasma flows, etc\cite{stangeby_plasma_2000}, or for other purposes
 \item Heavy impurity at trace levels (0.001\%), i.e.\,hydrogen and tungsten (W-H), where the tungsten usually originates from the walls and divertors made of tungsten.
\end{itemize}
The parameters chosen for these cases can be found summarized in Table \ref{table:values}. 
\begin{table}
\centering
 \begin{tabular}{|c||c|c|c|c|}
\hline \rule{0pt}{2ex}
     $\i-\j\rightarrow$      & \text{T-D} & \text{C-H} & \text{Ar-H} & \text{W-H} \\
        \hline\hline
 \rule{0pt}{3ex} $n_\i$   & $10^{19}$  &  $10^{18}$ & $10^{17}$ & $10^{14}$\\
  $Z_\i$   & $+1$       & $+6$       & $+7$      & $+7$\\ 
  $m_\i$   & $3$ amu    & $12$ amu   & $40$ amu  & $184$ amu\\
  $T_\i$   & $100$ eV   & $100$ eV   & $100$ eV  & $100$ eV\\   
  $n_\j$   & $10^{19}$  &  $10^{19}$ & $10^{19}$ & $10^{19}$\\
  $Z_\j$   & $+1$       & $+1$       & $+1$      & $+1$\\ 
  $m_\j$   & $2$ amu    & $1$ amu    & $1$ amu   & $1$ amu\\
  $T_\j$   & $50-200$ eV& $50-200$ eV& $50-200$ eV& $50-200$ eV\\
  \hline
\end{tabular}
\caption{Values of constants used for the different operational parameters. At 100eV, the maximum excitation state for higher-Z impurities is around +7, which the maximum charge states of Argon and Tungsten are limited to +7.}
\label{table:values}
\end{table}

\subsection{Convergence of transport coefficients}

For convergence calculations, we choose the maximum number of rank-1 and rank-2 moments to be $l_1=8$ and $l_2=8$ respectively, i.e. five and six more moments each in addition to $\mathbf{r}_{\i}$ and $\sigma_i$ respectively. The choice for this is mainly motivated by the compute time in Mathematica\cite{Mathematica}.

\begin{figure}
 \centering
 \includegraphics[width=\columnwidth]{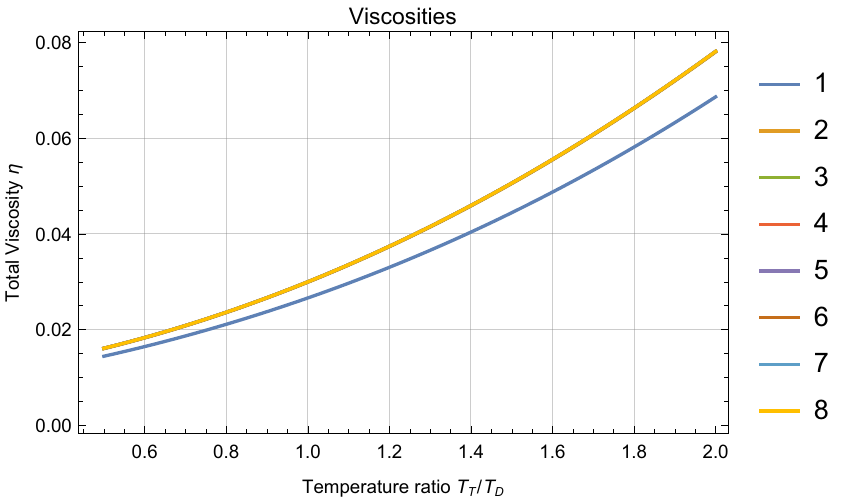}
 \caption{Plot of longitudinal viscosity w.r.t.\,the temperature ratio for multi-temperature $d_{\i\j}=\gamma_{\i\j}/2$ case on addition of multiple moments. Each number $M$ in the legend represents, in principle, an additional $5M$-moments. One can immediately notice that only the addition of $\sigma_\i$ as a moment contributes significantly, and every moment after that makes a much smaller contribution, so small in fact that the lines mostly seem to overlap beyond $M=2$.} 
 \label{fig:convergence_visc}
\end{figure}

\begin{figure}
 \centering
 \includegraphics[width=\columnwidth]{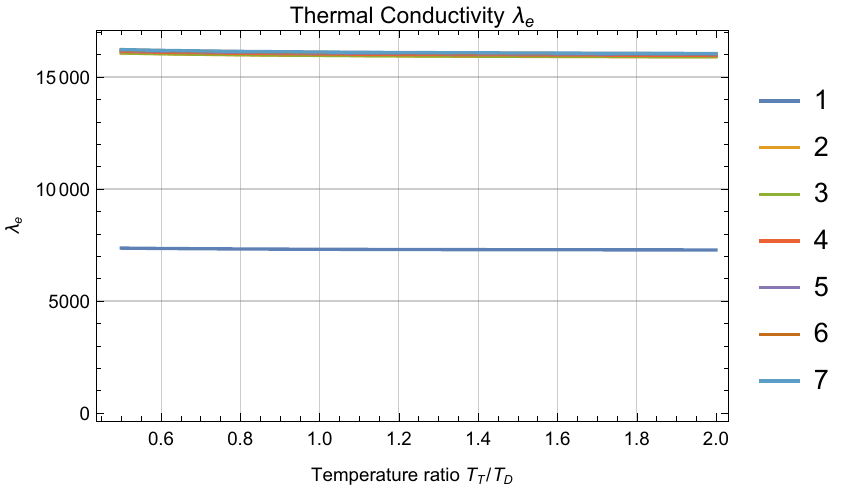}
 \includegraphics[width=\columnwidth]{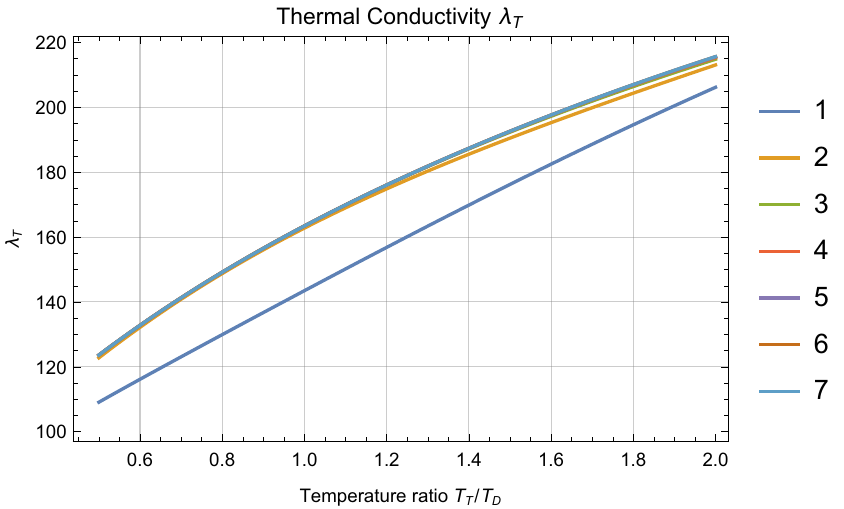}
 \includegraphics[width=\columnwidth]{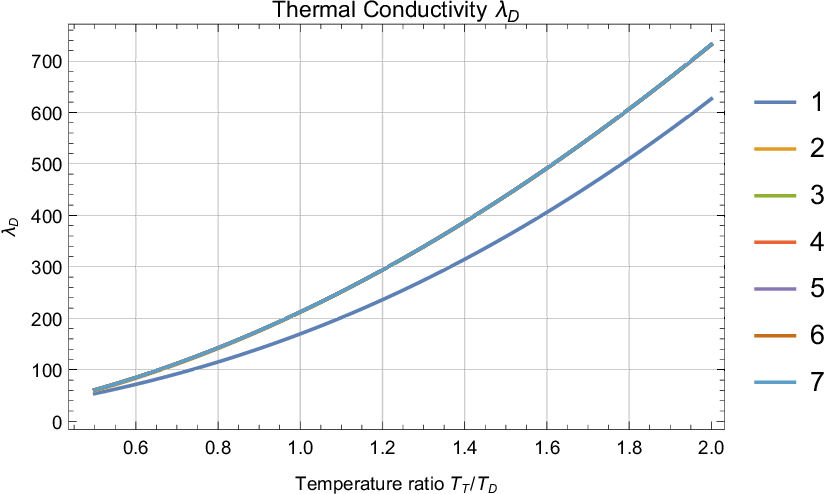}
  \caption{Plot of total parallel thermal conductivities $\lambda_\i$, for different species $\i$,for multi-temperature $d_{\i\j}=\gamma_{\i\j}/2$ case with the addition of more moments. The topmost plot is for $\lambda_e$, the middle one for $\lambda_D$ and the bottom one for $\lambda_T$. Each number in the legend represents, in principle, an addition of $3M$-moments. One can generally observe the same sort of convergence scheme as for the total longitudinal viscosity for the coefficients $\lambda_D$ and $\lambda_T$. Notice however the relatively slower convergence of the electron counterpart $\lambda_e$.}
  \label{fig:convergence_cond} 
\end{figure}

First, we graph the total longitudinal viscosity and total parallel heat conductivities for the D-T case against the temperature ratio for the multi-temperature coefficients with $d_{\i\j}=\gamma_{\i\j}/2$, and they can be found in Fig.\,(\ref{fig:convergence_visc}) and (\ref{fig:convergence_cond}) respectively.    
One can notice that the total longitudinal viscosity $\eta$ converges remarkably quickly. The addition of a second tensorial moment, i.e. $M=2$, results in a significant jump, but the addition of more moments simply seems to make the curves of the total longitudinal viscosity overlap.
Generally, we observe the same for the total parallel thermal conductivities, however, addition of a third moment in addition to $\mathbf{r}_\i$ seems to weakly contribute to the calculation of $\lambda_T$. This is in agreement with the results found by Balescu\cite{balescu_transport_1988}(Fig.\,4.1, pp.\,238), that addition of more moments beyond those in the $21N$-moment range seems to make no significant changes in these transport coefficients. Furthermore, the values of the longitudinal viscosity and parallel thermal conductivity values are well within the order of magnitude values as of Braginskii\cite{braginskii_transport_1965} (since they are all weakly proportional to $nkT\tau$ and $nk^2T\tau/m$ respectively, where $\tau$ is the mean collision time - see Ref.\,\onlinecite{raghunathan_generalized_2021} for details). The scaling between $\lambda_e$ and $\lambda_\alpha$ ($\alpha=D,H$) is also of the order of $(m_e/m_\i)^{1/2}$ as per Braginskii's estimations as well. A full comparison of further convergence and comparison with Braginskii's closure is a part of the planned future work. At this point, we have been able to verify that similar patterns follow for the generalized Zhdanov closure as observed previously in closures of Braginskii's and Balescu's type.

At this point, a question may be posed as to why we are plotting quantities with respect to the temperature ratio and not $\beta_\i=\omega_\i\tau_{\i\i}$ as has been done traditionally in Refs.\,\onlinecite{balescu_transport_1988,kaneko_transport_1960,ji_closure_2013}. In these references, a form of the collision operator is used which approximates the cross section such that the collision coefficients are essentially some rational multiple of $\tau_{\i\i}$, which is proportional to the Coulomb logarithm. Thus, on forming the matrix of collision coefficients and inverting the matrix, one essentially finds transport coefficients to be a polynomial of $\beta_\i$. However, as one can see from our collision cross section, there are other terms in addition to the logarithmic term, some of which may be significant compared to the Coulomb logarithm. This fact does not allow us to claim that our transport coefficients are merely polynomials of $\beta_\i$, making it an inconvenient plotting parameter. Furthermore, since we are only looking at the parallel/longitudinal component, the strength of the magnetic field is inconsequential to these parameters. We, therefore, choose the temperature ratio along with the distinction of the four physical cases, hoping to shed light on them individually. However, generalization of the multi-component parallel/longitudinal closure to magnetized closures will comprise a part of our future work, and we may address this problem again later in that context. 

In order to study these convergence effects more quantitatively, we calculate the differences from references values. The first is the maximum difference of the transport coefficients obtained with the multi-temperature coefficients with respect to the curve obtained for the same multi-temperature moments with maximum number of moments. We call this difference I, which allows us to see how quickly the transport coefficients from the multi-temperature scheme converge. The second difference II, is the maximum difference of the transport coefficients obtained with single-temperature coefficients with respect to the values obtained with the maximum number of moments using the single-temperature coefficients, i.e.\,$M=7$ for rank-1 coefficients and $M=8$ for rank-2 coefficients. This allows us to see how the transport coefficients in the single-temperature scheme converge. The third difference III we define is the difference of the single-temperature transport coefficients with respect to the values for the transport coefficients obtained from the multi-temperature scheme for the same number of moments, which allows us a global perspective of differences between the single and multi-temperature schemes. Now we calculate these difference values for the different transport coefficients using the multi-temperature and single-temperature coefficients for the four physical cases chosen, which can be found in Table \ref{table:errorsconvergence}. Again, to avoid difference values being exaggerated by transport coefficients approaching zero, we restrict the temperature ratio range to 0.5-2.
\begin{widetext}
\begin{center}
 \begin{table}
 \begin{tabular}{|c||c||ccc|ccc|ccc|ccc|}
   \hline
 \rule{0pt}{2ex} Coeff. & $M$ & \multicolumn{3}{c|}{D-T}  & \multicolumn{3}{c|}{H-C} & \multicolumn{3}{c|}{H-Ar} & \multicolumn{3}{c|}{H-W}\\\hline
 \rule{0pt}{2ex}& & I & II & III & I & II & III& I & II & III& I & II & III\\\hline 
 \multirow{7}{*}{$\eta$}& \rule{0pt}{2ex}1 & 12.309 & 13.061 & 30.873 & 36.596 & 36.767 & 67.057 & 23.324 & 21.115 & 68.269 & 14.287 & 11.316 & 67.958 \\
 &2 & 0.488 & 0.326 & 32.543 & 0.838 & 0.244 & 69.025 & 0.331 & 0.074 & 63.863 & 0.543 & 0.202 & 62.890 \\
 &3 & 0.043 & 0.012 & 32.337 & 0.021 & 0.008 & 67.967 & 0.042 & 0.042 & 63.448 & 0.028 & 0.011 & 62.295 \\
 &4 & 0.013 & 0.000 & 32.317 & 0.002 & 0.001 & 68.020 & 0.036 & 0.035 & 63.410 & 0.006 & 0.000 & 62.331 \\
 &5 & 0.002 & 0.000 & 32.318 & 0.000 & 0.000 & 68.015 & 0.026 & 0.018 & 63.433 & 0.002 & 0.000 & 62.332 \\
 &6 & 0.000 & 0.000 & 32.318 & 0.000 & 0.000 & 68.017 & 0.015 & 0.008 & 63.435 & 0.001 & 0.000 & 62.331 \\
 &7 & 0.000 & 0.000 & 32.318 & 0.000 & 0.000 & 68.016 & 0.006 & 0.003 & 63.440 & 0.001 & 0.000 & 62.331 \\\hline
 \multirow{7}{*}{$\lambda_e$}& \rule{0pt}{2ex}1 & 54.645 & 58.064 & 43.321 & 67.597 & 67.704 & 56.776 & 59.025 & 59.149 & 75.474 & 54.696 & 54.807 & 78.894 \\
 &2 & 0.929 & 1.094 & 54.750 & 0.563 & 0.567 & 57.293 & 0.810 & 0.813 & 76.028 & 0.930 & 0.933 & 79.400 \\
 &3 & 0.905 & 1.019 & 54.828 & 0.321 & 0.321 & 57.298 & 0.767 & 0.767 & 76.028 & 0.905 & 0.907 & 79.399 \\
 &4 & 0.515 & 0.570 & 54.921 & 0.157 & 0.157 & 57.298 & 0.419 & 0.418 & 76.034 & 0.515 & 0.515 & 79.408 \\
 &5 & 0.244 & 0.266 & 54.974 & 0.058 & 0.058 & 57.298 & 0.189 & 0.188 & 76.038 & 0.244 & 0.244 & 79.415 \\
 &6 & 0.090 & 0.097 & 54.998 & 0.016 & 0.016 & 57.298 & 0.066 & 0.066 & 76.041 & 0.090 & 0.090 & 79.420 \\\hline
 \multirow{7}{*}{$\lambda_\alpha$}& \rule{0pt}{2ex}1 & 20.076 & 21.509 & 42.780 & 60.105 & 58.795 & 77.898 & 38.690 & 34.508 & 71.104 & 20.248 & 14.950 & 66.987 \\
 &2 & 2.604 & 0.454 & 63.835 & 2.103 & 0.028 & 75.112 & 2.183 & 0.516 & 59.640 & 1.091 & 0.216 & 57.852 \\
 &3 & 0.102 & 0.053 & 60.142 & 0.147 & 0.001 & 71.224 & 0.539 & 0.450 & 59.456 & 0.080 & 0.017 & 56.311 \\
 &4 & 0.039 & 0.006 & 60.296 & 0.020 & 0.001 & 71.508 & 0.347 & 0.338 & 59.142 & 0.013 & 0.012 & 56.461 \\
 &5 & 0.007 & 0.004 & 60.303 & 0.004 & 0.001 & 71.473 & 0.219 & 0.187 & 59.347 & 0.007 & 0.007 & 56.464 \\
 &6 & 0.003 & 0.002 & 60.294 & 0.013 & 0.001 & 71.497 & 0.100 & 0.074 & 59.363 & 0.004 & 0.003 & 56.461 \\\hline
 \multirow{7}{*}{$\lambda_\beta$}& \rule{0pt}{2ex}1 & 12.930 & 13.180 & 19.553 & 156.325 & 85.818 & 26.297 & 100.557 & 101.934 & 34.412 & 17.412 & 17.468 & 68.823 \\
 &2 & 1.274 & 0.498 & 21.753 & 23.692 & 2.043 & 37.972 & 35.468 & 21.273 & 36.624 & 4.529 & 3.400 & 70.707 \\
 &3 & 0.372 & 0.009 & 21.579 & 8.199 & 0.045 & 60.988 & 21.158 & 17.797 & 31.046 & 1.177 & 0.995 & 67.853 \\
 &4 & 0.144 & 0.005 & 21.499 & 4.973 & 0.510 & 66.775 & 14.944 & 13.116 & 29.805 & 0.295 & 0.014 & 68.310 \\
 &5 & 0.033 & 0.002 & 21.493 & 2.718 & 0.137 & 69.818 & 9.673 & 7.875 & 27.889 & 0.146 & 0.131 & 68.571 \\
 &6 & 0.007 & 0.001 & 21.493 & 0.999 & 0.048 & 72.413 & 4.336 & 3.463 & 25.724 & 0.061 & 0.062 & 68.613 \\\hline 
 \end{tabular}
 \caption{Values of percentage differences of different types, by the addition of more moments, for the four different physical cases. The table is divided into rows, firstly by the transport coefficient, and then by the number of moments, and it is divided into columns first by the physical case in question, and then the type of the difference in question. Appropriate boundaries have been drawn for ease of reading.}
 \label{table:errorsconvergence}
 \end{table} 
\end{center}
\end{widetext}
From the table, a few observations can be made
\begin{enumerate}
\item The most significant difference seems to indeed remain between the $13N$-moment scheme and the others. An improvement in the range of $10-150\%$ can be seen, with the lighter impurities representing the lower end of the range, and the mid-weight impurities representing the higher end. 
 \item The total longitudinal viscosities seem to converge much faster than the total thermal conductivities in general. All viscosities seem to converge to two decimal places in just the second moment (i.e.\,the $21N$-moment scheme). 
 \item In general, the differences for any transport coefficient seem to increase as the weight of the impurity increases and attain a maximum between the Carbon and Argon cases. 
 \item The thermal conductivity $\lambda_\j$ ($\j=T,C,Ar,W$), seems to not even converge to one decimal place for Carbon and Argon within the $21N$-moment scheme. The thermal conductivity $\lambda_\beta$ ($\j=T,C,Ar,W$), seems to converge better than $\lambda_\i$, but still exhibits significant differences for Carbon and Argon. Additionally, one can roughly say that the thermal conductivities converge to one decimal place in the $21N$-moment scheme are not converged to one decimal place for mid-weight impurities. 
 \item The thermal conductivity $\lambda_e$ seems to converge to two decimal places in the $21N$-moment scheme even for mid-weight impurities, in contrast with the other two thermal conductivities $\lambda_\i$ and $\lambda_\j$. Furthermore, it seems to converge to two decimal places in the scope of the $21N$-moment scheme, slightly better than the latter conductivities. This seems to corroborate why extension of the $13N$-moment scheme to the $21N$-moment scheme was considered sufficient for the $\lambda_e$ in previous works.  
 \item Differences of types I and II seem to more or less be of the same order for any considered moment. 
 \item From the type III columns, we can observe that the differences between the single-temperature scheme and the multi-temperature scheme remain quite significant at any number of moments, and remain virtually the same beyond the $21N$-moment scheme. The average difference between these two schemes seems to be in the range of $10-80\%$, with the mid-weight impurities representing the higher side of this range. Furthermore, even though it might seem as if they type II errors are converging marginally faster for the mid-weight impurities, significant differences still remain compared to the multi-temperature scheme (type III), even for heavy impurities. 
 Based on this, at least for mid-weight impurities and heavy impurities, we recommend using multi-temperature coefficients. 
\end{enumerate}

\subsection{Convergence of the friction and thermal forces}

A corollary, but important aspect to the effect of addition of more moments on the transport coefficients, is its effect on the finally calculated values of the friction and temperature-gradient dependent forces in the RHS of the momentum equations. This is especially so, because the fluid code packages mentioned in the previous section explicitly use the Zhdanov closure scheme to calculate their friction and thermal forces, making it necessary to study the implementation of the scheme. The augmentation of the friction force can be studied, for any moment with respect to the $5N$-moment scheme, by the following expression
\begin{equation}
 \%_{fric,aug}(l_1)=\frac{M^{1}_{l_1 0}(M^{1}_{l_1})^{-1}M^{1}_{0l_1}}{M^{100}}\times 100\%,\label{eq:fric_aug}
\end{equation}
where, since the $M$-matrices are $N\times N$ in dimension, the division is performed element wise. Here we use the term ``augmentation'' in a general sense - additional moments may also contribute to reducing the values of the forces. We studied the augmentation of the friction force for the case of $l_1=1$ in the previous article\cite{raghunathan_generalized_2021}. The addition of the thermal force can be defined with respect to the first contribution from the heat-flux terms., i.e.\,w.r.t.\,the $13N$-moment scheme (since there is no temperature gradient force in the $5N$-moment scheme), as follows
\begin{equation}
 \%_{therm,aug}(l_1)=\frac{M^{1}_{l_1 0}(M^{1}_{l_1})^{-1}\Lambda^{1}_{l_1 }-M^{1}_{1 0}(M^{1}_{1})^{-1}\Lambda^{1}_{1 }}{M^{1}_{1 0}(M^{1}_{1})^{-1}\Lambda^{1}_{1 }}\times 100\%.\label{eq:therm_aug}
\end{equation}
One element in these matrices of special interest, is the effect of the background flow on the impurity, which in this case is the effect of deuterium flow on the tritium friction force term, since this term is what affects the impurity dynamics.

\begin{figure*}
\centering
 \includegraphics[width=\columnwidth]{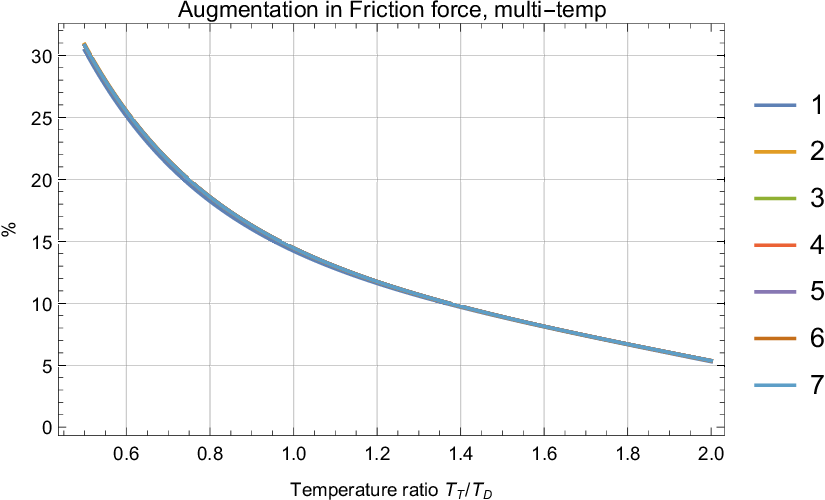}\includegraphics[width=\columnwidth]{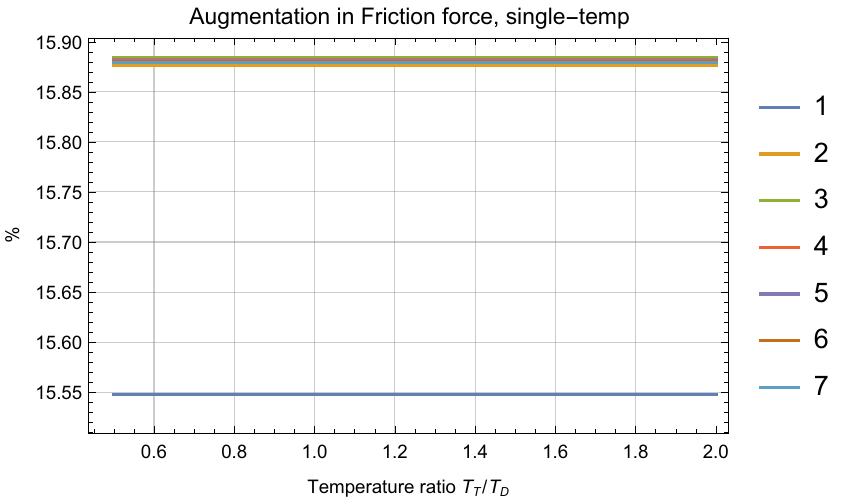}
 \caption{Plot of augmentation percentage for friction forces for the multi-temperature (top) and the single-temperature (bottom) cases, on addition of moments. One can notice that the curves beyond the addition of the $13N$-moment case are more-or-less overlapping. One can notice that the single-temperature coefficients do not suggest much variation in the augmentation over the temperature range.}
 \label{fig:friction_forces}
\end{figure*}

\begin{figure*}
\centering
 \includegraphics[width=\columnwidth]{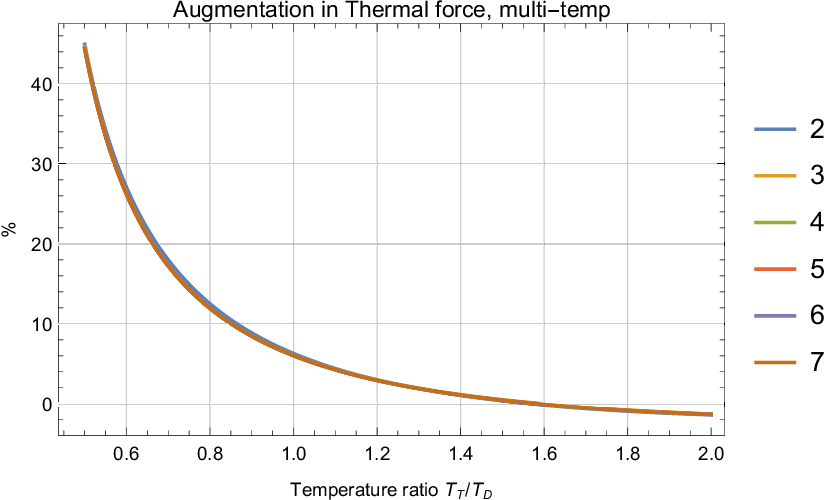}\includegraphics[width=\columnwidth]{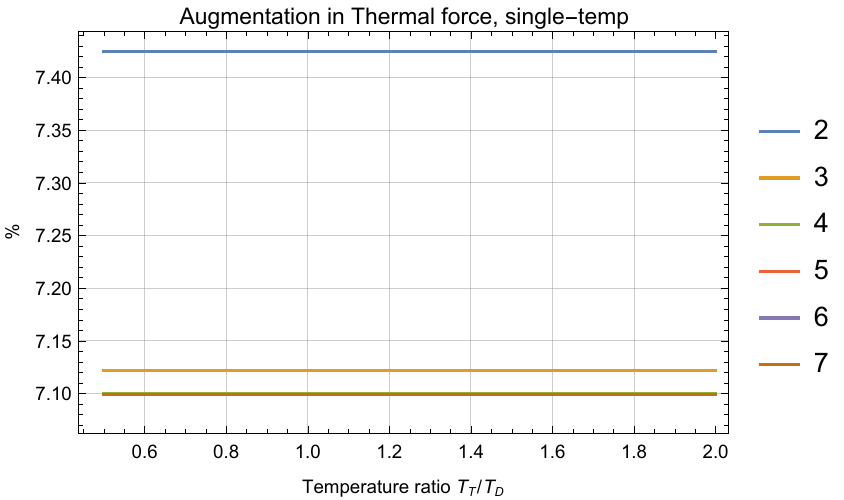}
 \caption{Plot of thermal force augmentation percentages for the multi-temperature (top) and the single-temperature (bottom) cases, on addition of moments. One can notice that the curves beyond the addition of the $13N$-moment case are more-or-less overlapping. Similar to the case for the friction forces, the single-temperature coefficients do not suggest much variation in the augmentation of thermal forces over the temperature range.}
 \label{fig:thermalforces}
\end{figure*}

We first proceed to plot, as in the previous sections, the augmentations of the friction and thermal forces for the deuterium-tritium case for this element, in Figs.\,\ref{fig:friction_forces} and \ref{fig:thermalforces} respectively. 
One can see that the forces are converged relatively quickly on the addition of just the second moment, i.e. under the $21N$-moment scheme. However, a key difference between the single-temperature and multi-temperature scheme is that the single-temperature scheme suggests very little variation in friction forces over the temperature range, whereas the multi-temperature scheme suggests a much larger range of magnitude of friction forces over the chosen temperature range. 
A similar observation can be made for the thermal forces as well. Generally, the friction and thermal forces for the deuterium-tritium case seem converged adequately near the unity temperature ratio for the $21N$-moment case, but diverge as the temperature ratio goes further away from one.

Next, as in the previous subsection, we form tables of augmentation of the friction and thermal forces as defined in Eqs.\,(\ref{eq:fric_aug}) and (\ref{eq:therm_aug}) for the four different physical cases chosen. As before, we define three types of augmentations. Type I refers to the augmentations for the multi-temperature case, Type II represents augmentations for the single-temperature case, and Type III denotes the percentage difference of the single-temperature force w.r.t.\,the multi-temperature force for any given moment. As mentioned earlier, we consider the force coefficients for the forces on the impurity due to the main-ion flow velocity. The tables for the friction force augmentation and the the thermal force augmentation can be found in Tables \ref{tab:fric} and \ref{tab:therm}.
\begin{widetext}
\begin{center}
 \begin{table}
 \begin{tabular}{|c||ccc|ccc|ccc|ccc|}
   \hline
 \rule{0pt}{2ex} $M$ & \multicolumn{3}{c|}{D-T}  & \multicolumn{3}{c|}{H-C} & \multicolumn{3}{c|}{H-Ar} & \multicolumn{3}{c|}{H-W}\\\hline
 \rule{0pt}{2ex} & I & II & III & I & II & III& I & II & III& I & II & III\\\hline
  \rule{0pt}{2ex} 1 & 30.396 & 15.547 & 47.759 & 48.639 & 48.421 & 67.327 & 29.996 & 27.274 & 65.231 & 14.453 & 0.997 & 74.049 \\
 2 & 30.892 & 15.876 & 48.073 & 48.679 & 48.423 & 66.654 & 30.302 & 28.538 & 62.008 & 13.723 & 1.001 & 72.570 \\
 3 & 30.800 & 15.884 & 48.094 & 48.679 & 48.424 & 66.636 & 30.903 & 29.098 & 61.453 & 13.372 & 1.005 & 71.864 \\
 4 & 30.785 & 15.882 & 48.084 & 48.680 & 48.424 & 66.637 & 31.060 & 29.389 & 61.315 & 13.376 & 1.005 & 71.871 \\
 5 & 30.787 & 15.881 & 48.079 & 48.681 & 48.424 & 66.637 & 31.171 & 29.550 & 61.218 & 13.427 & 1.005 & 71.972 \\
 6 & 30.786 & 15.880 & 48.076 & 48.683 & 48.425 & 66.638 & 31.222 & 29.647 & 61.132 & 13.449 & 1.005 & 72.017 \\
 7 & 30.786 & 15.879 & 48.076 & 48.685 & 48.426 & 66.639 & 31.257 & 29.708 & 61.066 & 13.451 & 1.004 & 72.022 \\\hline
 \end{tabular}
 \caption{Differences in the augmentation of the friction force on addition of more moments.}
 \label{tab:fric}
 \end{table}
  \begin{table}
 \begin{tabular}{|c||ccc|ccc|ccc|ccc|}
   \hline
 \rule{0pt}{2ex} $M$ & \multicolumn{3}{c|}{D-T}  & \multicolumn{3}{c|}{H-C} & \multicolumn{3}{c|}{H-Ar} & \multicolumn{3}{c|}{H-W}\\\hline
 \rule{0pt}{2ex} & I & II & III & I & II & III& I & II & III& I & II & III\\\hline
  \rule{0pt}{2ex}2 & 44.953 & 7.425 & 68.190 & 5.283 & 0.819 & 52.278 & 15.449 & 15.318 & 40.846 & 17.836 & 15.233 & 41.451 \\
 3 & 44.665 & 7.122 & 68.051 & 5.738 & 0.808 & 53.029 & 16.560 & 15.761 & 46.712 & 13.880 & 13.899 & 48.659 \\
 4 & 44.350 & 7.100 & 68.382 & 5.769 & 0.808 & 53.080 & 17.037 & 16.193 & 47.433 & 14.104 & 13.448 & 49.461 \\
 5 & 44.465 & 7.099 & 68.248 & 5.768 & 0.806 & 53.082 & 17.315 & 16.563 & 47.283 & 13.792 & 13.403 & 48.856 \\
 6 & 44.443 & 7.099 & 68.273 & 5.768 & 0.804 & 53.084 & 17.588 & 16.810 & 47.289 & 13.869 & 13.424 & 48.524 \\
 7 & 44.442 & 7.099 & 68.275 & 5.766 & 0.804 & 53.081 & 17.808 & 16.969 & 47.407 & 13.852 & 13.441 & 48.471 \\\hline
 \end{tabular}
 \caption{Differences in the maximum augmentation of the thermal force on addition of more moments.}
 \label{tab:therm}
 \end{table}
 \end{center}
\end{widetext}
From the table of the friction forces, we can generally observe that irrespective of the single-temperature or multi-temperature coefficients, the friction force seems to converge to two decimal places within the $21N$-moment approximation compared to the friction force of the $5N$-scheme. In fact, the $13N$-moment scheme seems to increase the friction force most significantly, with the $21N$-moment scheme delivering further precision. We can also observe that there still remain significant differences between the single-temperature and multi-temperature friction forces at any given number of moments, which does not decrease significantly on addition of more moments. They also seem to remain more-or-less of the same order for increasing impurity weight. However, a part of this effect is counteracted by the fact that the augmentations themselves seem to be very small for heavy impurity case of tungsten, about 1\% or less. However, for  mid-weight impurities such as carbon and argon, using multi-temperature coefficients may be necessary for more precise representation of the friction force, as the augmentations to the friction force remain significant. 

\begin{figure}
\centering
 \includegraphics[width=\columnwidth]{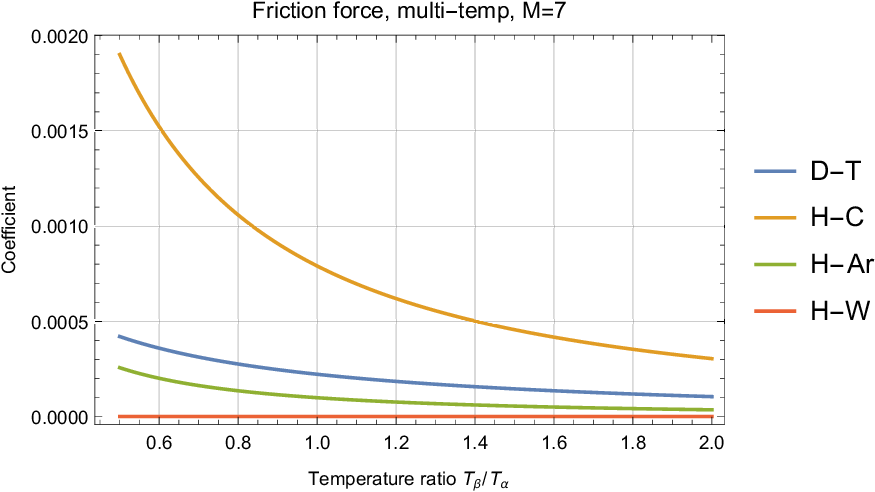}
 \includegraphics[width=\columnwidth]{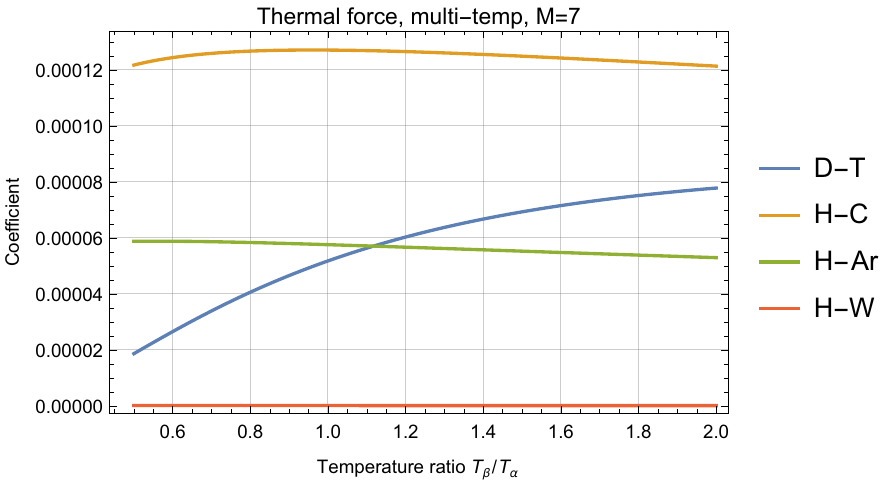}
 \caption{Plot of friction and thermal force coefficients, i.e. the coefficients of the flow velocities and the temperature gradients in the RHS of the momentum equation, for the multi-temperature case, for $M=7$, i.e. seven additional moments}
 \label{fig:friction_forces3} 
\end{figure}

In mild contrast, the thermal forces generally seem to converge to one decimal place or less within the $21N$-moment scheme, however, the addition by the $21N$-moment scheme is significant for low-weight and mid-weight impurities. For the deuterium-tritium and carbon case, the $21N$-moment thermal force augmentation seems fairly close to the values on addition of further moments. However, for argon and tungsten, the convergence of the augmentations of friction and thermal forces is much slower. For tungsten however, the friction and  thermal forces themselves remain much smaller compared to mid-weight impurities in higher concentrations, as can be noticed from Fig.\,\ref{fig:friction_forces3}, therefore, a slow convergence does not affect the transport coefficients related to it significantly. Furthermore, similar differences between the multi-temperature and single temperature coefficients, as seen from the Type III difference, remain significant. (The ones for the D-T case are slightly exaggerated by thermal force being close to zero for a certain temperature ratio. But from Fig.\,(\ref{fig:friction_forces}), one can see that the differences are still significant). 

On the basis of the studies on the friction force and the thermal force, we find generally that the $21N$-moment scheme brings added precision to the force values over the $13N$-moment scheme. Again however, we advise caution while using the single-temperature coefficients for mid-weight impurities, since the convergence of the forces is much lesser than that of low-weight impurities, and since the force augmentation is much more significant than heavy impurities. 

\subsection{Section summary}

Based on the observations in the two previous subsections, we can generally say that the $21N$-moment system represents a significant gain in precision of transport coefficients over the $13N$-system. More precisely, the $21N$-moment system can represent the total viscosity very well (to two decimal places), but the total thermal conductivities are represented only moderately well (to maximum one decimal place). Furthermore, there still persist significant differences between the single-temperature coefficients and multi-temperature coefficients, particularly for mid-weight impurities and heavy impurities. Consequently, even the converged values of transport coefficients for multi-temperature and single-temperature coefficients differ significantly. Similar observations are made for the convergence of the friction forces and the thermal forces are made as well, with the friction force being precise to two decimal places in the $21N$-moment scheme, and the thermal forces to one decimal place, with similar reservations about the convergence for mid-weight impurities. 
Thus, caution is recommended when using the $21N$-moment single-temperature scheme for mid-weight impurities such as Carbon and Argon, and heavy impurities such as tungsten. For them, the multi-temperature scheme may be better suited. We have thus, with the help of the convergence study, managed to establish the significant improvement brought over by the multi-temperature $21N$-moment scheme, and have also managed to establish a range of validity for the existing $21N$-moment single-temperature which comprise the Zhdanov closure.

\section{The $21N$-moment Zhdanov closure scheme vs trace approximation}
\label{sec:21n}

At this juncture, it would be interesting to study the performance of trace approximation schemes with respect to $21N$-moment Zhdanov closure, because the earlier versions of some of the fluid code packages often depended on some sort of trace approximation on the impurities for simulations.
Having demonstrated in the previous section that the $21N$-moment schemes bring the values of transport coefficients, and, friction and thermal forces to two decimal places of convergence, we can use it as a standard to study the accuracy of any trace approximation scheme. The advantage of using trace approximation in the Zhdanov closure scheme is that they are computationally lighter as compared to the inclusion of all collisional coefficients. This is an important consideration, as generally in the numerical implementation of the Zhdanov closure, the calculation of the collision coefficients is what takes the most amount of time, the inversion of the matrices, e.g.\,in Eqs.\,(\ref{eq:generalizedrank1}) and (\ref{eq:generalizedrank2}), being relatively quick\cite{bufferand_2019}. Thus, if the domain of validity of the trace approximation is established, one could identify certain scenarios in which the fluid code could switch to the trace approximation to save compute time.

In order to apply the trace approximation in the Zhdanov closure, the impurities are assumed to be at such low concentrations that they have no effect on the background plasma at work. Thus, one way to define the trace approximation as follows
\begin{enumerate}
\item The self-collisions for impurities are excluded by the means of excluding the $(A_{\i\i}^{mnl}+B_{\i\i}^{mnl})$ term in Eq.\,(\ref{eq:mmnl}) in $M^{mnl}_{\i\j}$. 
\item Terms $A_{\i\j}^{mnl}$, $B_{\i\j}^{mnl}$ are neglected for electron-impurity collisions, i.e. we neglect any effects collisions with impurities have on electrons.
\item Furthermore, we neglect $A_{\i\j}^{mnl}$, $B_{\i\j}^{mnl}$ for the ion-impurity collisions, i.e. we assume the impurities are not in a sufficient quantity to affect the main ion species.
\end{enumerate}
Then this trace approximation assumptions are used for the collision coefficients which enter the $21N$-moment Zhdanov closure, and the transport coefficients and the friction/thermal forces are calculated in that manner. In principle, about 33\% of compute time can be saved by ignoring these three coefficients.

On comparing the transport coefficients and the friction and thermal forces for the multi-temperature trace tungsten case, we find that the deviations (excepting for $\lambda_W$, which generally has a negligible contribution to the heat-flux for trace density values) are less than 1\% and 0.1\% respectively. This is expected, as such trace approximations are usually only applicable to impurities at very low concentrations, such as tungsten being present at $10^{-5}$ the density of the main ion species. They are generally valid in either fluid codes where there is some uncertainty of values of transport coefficients between impurities and the main species, when on doing the trace approximation at least would ensure no spurious contributions from the impurity to the main plasma. They are also used in some orbit-following codes where it is hard to quantify the back-reaction of the impurities on the plasma, because the main plasma is taken to be effectively static. Generally, such codes are unable to simulate impurities at significant concentrations. 

Thus, it is worth studying the transport coefficients for the three other physical cases mentioned relevant to SOL/edge plasmas as in the previous section. As in the previous section, we calculate and tabulate the differences from a reference curve. We can safely take the the $21N$-moment, multi-temperature values with full collisions included, for each physical case, as the reference, and define differences with respect to this curve, defined as follows
\begin{equation}
 \%_{diff,\ case}= \left|\frac{t_{case}-t_{21N,multi-temp,full}}{t_{21N,multi- temp,full}}\right|\times 100\%.
 \label{eq:percentagediff}
\end{equation}
The of the maximum percentage errors for different cases can be found in Table \ref{table:errors21n}.
\begin{center}
 \begin{table}
  \begin{tabular}{|c||c|c|c|}
  \hline
  \multicolumn{1}{|c||}{\rule{0pt}{2ex}}&\multicolumn{1}{c|}{Full collisions}&\multicolumn{2}{c|}{Trace Approx}\\\hline
  \multicolumn{1}{|c||}{\rule{0pt}{2ex}}&\multicolumn{1}{c|}{Single-temp}&\multicolumn{1}{c|}{Mutli-temp}&\multicolumn{1}{c|}{Single-temp}\\\hline
   \rule{0pt}{2ex}$\eta$&   23.684 & 120.481 & 114.106 \\
   $\lambda_e$ &             37.895 & 21.916 & 68.089 \\
   $\lambda_T$ &             10.177 & 34.256 & 32.807 \\
    $\lambda_D$ &            47.970 & 121.388 & 227.292 \\\hline
    \rule{0pt}{2ex}$\eta$&   67.572 & 265.618 & 506.460 \\
    $\lambda_e$ &             57.293 & 71.213 & 169.382 \\
    $\lambda_C$ &             37.972 & 474.925 & 288.010 \\
    $\lambda_H$ &             75.112 & 214.212 & 418.211 \\\hline
   \rule{0pt}{2ex}$\eta$&    63.385 & 39.183 & 126.947 \\
   $\lambda_e$ &              76.028 & 13.333 & 99.526 \\
   $\lambda_Ar$ &             36.624 & 36.673 & 28.776 \\
   $\lambda_H$ &              59.640 & 38.127 & 120.113 \\\hline
   \rule{0pt}{2ex}$\eta$&  61.988 & 0.513 & 62.055 \\
   $\lambda_e$ &            79.400 & 0.014 & 79.426 \\
   $\lambda_W$ &            70.707 & 37.456 & 53.905 \\
   $\lambda_H$ &            57.852 & 0.049 & 57.930 \\\hline
  \end{tabular}
\caption{Values of maximum percentage differences from the values of the transport coefficients from the ones for the $21N$-moment multi-temperature full collisions case, as defined in Eq.\,(\ref{eq:percentagediff}). The temperature ratio range is limited between 0.5-2, because near for some coefficients, for smaller values of temperature ratio, some of the transport coefficients tend to approach zero very quickly, exaggerating the relative differences.}
\label{table:errors21n}
 \end{table}
 \end{center}
From this table, we can find a few general patterns. Firstly, the differences in transport coefficients seem to increase initially with impurity weight/charge and then decrease. This is consistent with our observations in our previous article\cite{raghunathan_generalized_2021}, that the differences attain a maxima in between the charges/densities of Carbon and Argon. This is partly consistent with increasing differences observed by Balescu et al\cite{balescu_transport_1988} with increasing charge state $Z_\i$, who did not observe a decrease because they did not consider the densities to decrease as the charge state increased. Thus, the decrease observed in our case arises mainly from the decreasing densities across our four physical cases.
And second, the transport coefficients calculated from single-temperature coefficients exhibit larger differences from those calculated from their multi-temperature counterparts.

Concerning the friction and thermal forces, as in the previous section, we compare the coefficient of the force on the impurity due to the background ion flow or temperature gradient.  However, for low and mid-weight impurities, we find that though friction forces for all cases lie within $30\%$ of each other, the thermal forces are overestimated by the trace approximation by a factor of two. The maximum difference is observed for the carbon case, similar to what was observed earlier for the case of transport coefficients, as can be seen in Tables \ref{table:errorsforces21n} and \ref{table:errorsthermalforces21n}. 
We can also notice that the single-temperature coefficients present significant differences compared to the multi-temperature ones, in agreement with all our observations so far. 
\begin{center}
 \begin{table}
  \begin{tabular}{|c||c|c|c|}
  \hline
  \multicolumn{1}{|c||}{\rule{0pt}{2ex}}&\multicolumn{1}{c|}{Full collisions}&\multicolumn{2}{c|}{Trace Approx}\\\hline
  \multicolumn{1}{|c||}{\rule{0pt}{2ex}}&\multicolumn{1}{c|}{Single-temp}&\multicolumn{1}{c|}{Mutli-temp}&\multicolumn{1}{c|}{Single-temp}\\\hline
   \rule{0pt}{2ex}D-T&  54.940 & 18.538 & 43.400 \\
   H-C &             66.654 & 44.081 & 63.143\\
   H-Ar &            62.008 & 28.619 & 56.706 \\
    H-W &           72.570 & 0.077 & 72.464 \\\hline
  \end{tabular}
\caption{Maximum percentage difference in friction force compared to the full $21N$-moment full collisional values for different cases.}
\label{table:errorsforces21n}
 \end{table}
 \end{center} 
 \begin{center}
 \begin{table}
  \begin{tabular}{|c||c|c|c|}
  \hline
  \multicolumn{1}{|c||}{\rule{0pt}{2ex}}&\multicolumn{1}{c|}{Full collisions}&\multicolumn{2}{c|}{Trace Approx}\\\hline
  \multicolumn{1}{|c||}{\rule{0pt}{2ex}}&\multicolumn{1}{c|}{Single-temp}&\multicolumn{1}{c|}{Mutli-temp}&\multicolumn{1}{c|}{Single-temp}\\\hline
   \rule{0pt}{2ex}D-T&   58.273 & 126.168 & 219.337\\
   H-C &            52.278 & 460.064 & 733.785 \\
   H-Ar &            40.846 & 81.030 & 154.847 \\ 
    H-W &            41.451 & 0.089 & 41.578 \\\hline
  \end{tabular}
\caption{Maximum percentage difference in thermal force compared to the full $21N$-moment full collisional values for different cases.}
\label{table:errorsthermalforces21n}
 \end{table}
 \end{center}

\begin{figure*}
 \centering
 \includegraphics[width=\columnwidth,trim= 0 50 0 0,clip]{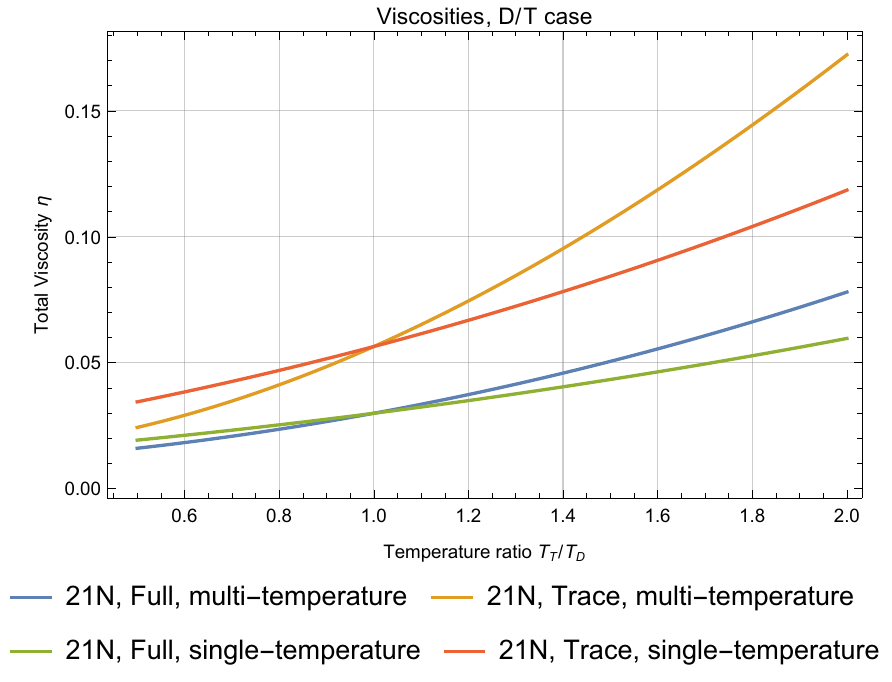}\includegraphics[width=\columnwidth]{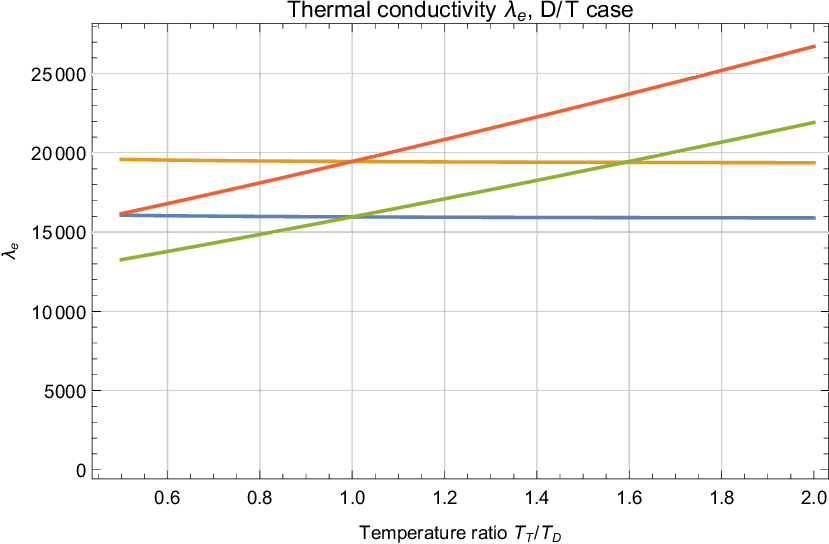}
 \includegraphics[width=\columnwidth]{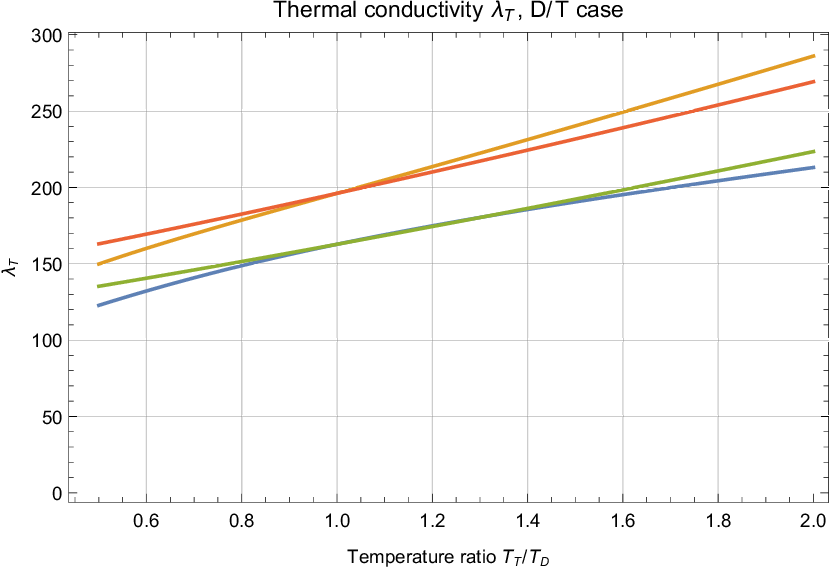}\includegraphics[width=\columnwidth]{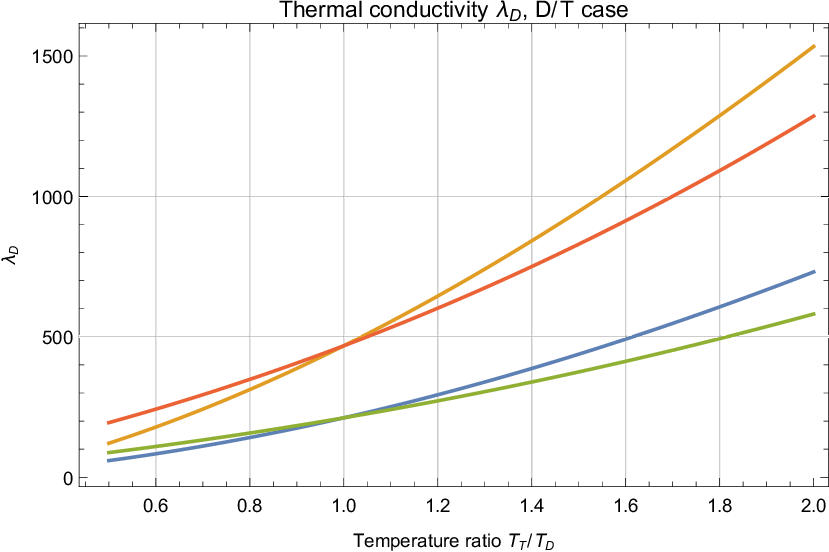}
 \includegraphics[trim= 0 0 0 270,clip,width=\columnwidth]{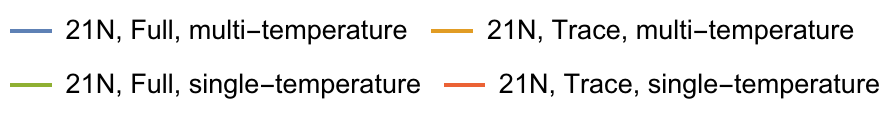}
  \caption{Plot of total longitudinal viscosity $\eta$ and total parallel thermal conductivities $\lambda_\i$, for different species $\i$,for multi-temperature and single-temperature case, for the case of full collisions and trace approximation, for the $13N$-moment and $21N$-moment approximations, with respect to the temperature ratio. The curves the cool and warm colours represent the cases of full collisions and trace-approximation respectively.}
  \label{fig:21nconductivitiesdt_all}
\end{figure*}

\begin{figure*}
\centering
 \includegraphics[width=\columnwidth]{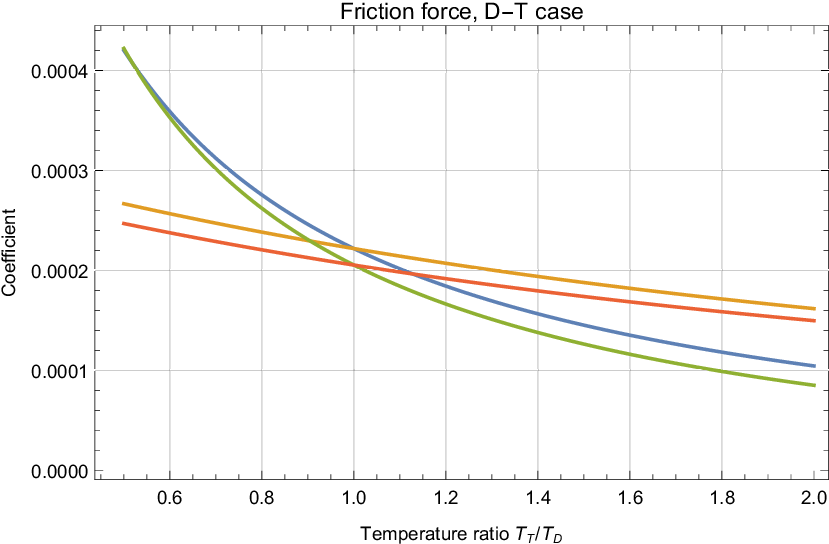}\includegraphics[width=\columnwidth]{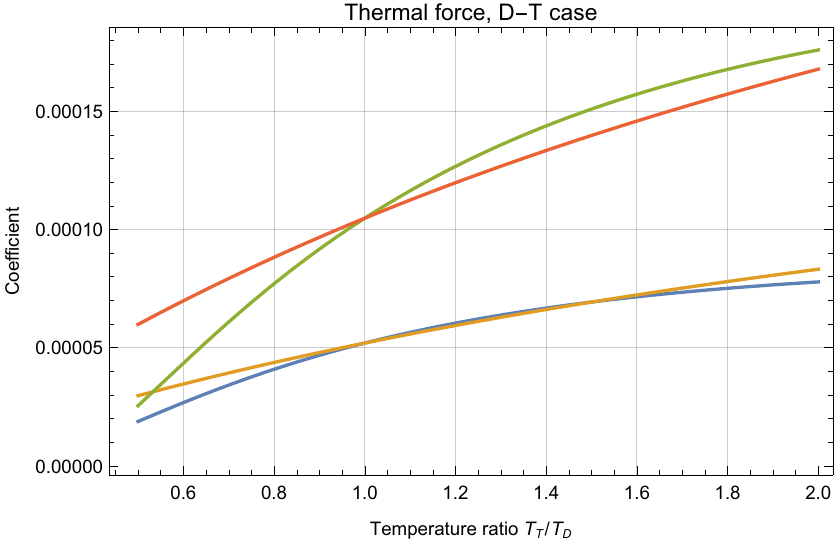}
  \includegraphics[width=\columnwidth,trim= 0 0 0 270,clip]{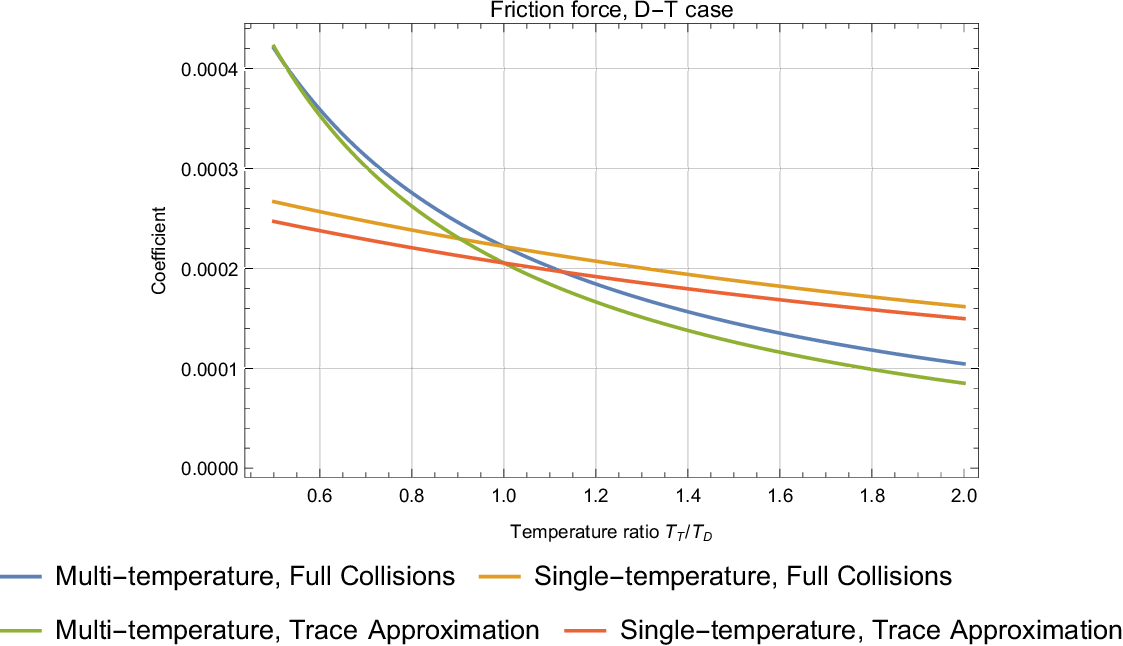}
 \caption{Friction and Thermal force coefficients for the D-T case. } 
 \label{fig:dttrace}
\end{figure*}

In order to illustrate better the differences of using the trace approximation in a case which does not allow for it, we proceed to plot the values of the transport coefficients for multi-temperature and single-temperature cases, for the full collisions and trace approximation cases. Fig.\,\ref{fig:21nconductivitiesdt_all} represents the plot of the total longitudinal viscosity and total parallel thermal conductivities, for the trace approximation and full closure, for the D-T case. 
From the plot for viscosity and thermal conductivities, one can indeed notice that the trace approximation overestimates the transport coefficients significantly. Furthermore, the curves with the single-temperature and multi-temperature schemes still follow each other quite closely, only separating out significantly for trace approximation. The thermal conductivities seem to be closer to each other, with a difference of about $30-50\%$ between the trace approximation and full collisions, but the viscosity is overestimated by a factor of two nearly. 
One can also notice from Fig.\,\ref{fig:dttrace}, that indeed though the friction forces follow each other to some 30\% difference, the thermal force is overestimated by a factor of two nearly.

It is worth mentioning that there is also a heuristic manner in which the trace approximation can be formulated, which lends itself to a much faster computational scheme than the abovementioned full and trace Zhdanov closure. This involves the same assumptions on the collision coefficients, but then proceeds to formulate the friction and thermal forces on the particles by the means of solving the thermal diffusion of impurities from the momentum equation at slow time evolution and small spatial gradients approximation\cite{stangeby_plasma_2000,chapman_thermal_1958,rutherford_impurity_1974}, with the collision coefficients at common plasma temperature. Generally, this procedure gives a heuristic estimate on the coefficients of the temperature gradients in the thermal force, where both coefficients are generally proportional to the square of the impurity charge state  $Z_\i^2$, and for heavy impurities, the ion thermal gradient coefficient is of the same order to the electron one, such that their ratio is of order $O(1)$. If the thermal force is such that 
\begin{equation}
 F_{therm\parallel}\sim\alpha_e\nabla_{\parallel} T_e + \beta_i\nabla_{\parallel} T_i, 
\end{equation}
then the values of $\alpha_i$ and $\beta_e$ are given by
\begin{align}
 \i_e&=\i_{e0} Z^2\\
 \j_i&=3\frac{\mu+5\sqrt{2}Z^2(1.1\mu^{5/2}-0.35\mu^{3/2})-1}{2.6-2\mu+5.4\mu^2}
\end{align}
where $\mu=m_Z/(m_i+m_Z)$ the relative mass ratio of the impurity to the main ion, and $Z$ is the charge state of the impurity. This heuristic scheme can also be considered as a formulation of the thermal force in the static $5N$-moment single-temperature approximation, which is also the fastest in terms of performance. According to Ref.\,\onlinecite{stangeby_plasma_2000,rutherford_impurity_1974}, $\i_{e0}=0.71$, and according to Ref.\,\onlinecite{chapman_thermal_1958}, $\i_e=1.0$. Thus, a range of heuristic thermal force scaling factors $\j_i/\i_e$ can be defined of the form
\begin{equation}
 (3,4.23)\times\frac{\mu+5\sqrt{2}Z^2(1.1\mu^{5/2}-0.35\mu^{3/2})-1}{(2.6-2\mu+5.4\mu^2)Z^2},
\end{equation}
where this scaling factor measures the relative magnitude of the forces from the ion and electron temperature gradients. 
We can state that for the tungsten case at unity temperature ratio, the scaling is followed very well, as is expected at the trace limit. But there exist significant deviations outside the vicinity of equal temperatures. Furthermore, there are deviations for mid-weight impurities, where this heuristic approximation overestimates the scaling compared to Zhdanov closure in the trace limit. Therefore, such a heuristic thermal force scheme is not recommended for mid-weight impurities at significant densities. This scheme is only strictly valid for the trace impurity case, and where the temperatures of all species are same, which it were to be the case would be the least computationally expensive. 

Thus, from the observations in this section, we can state confidently that the trace approximation by neglecting certain collisional coefficients is indeed valid for the trace heavy impurity case, while exhibiting significant differences in the transport coefficients and the thermal force for the mid-weight impurities at significant density fraction. 
Thus, on this basis, we can recommend safely replacing the $21N$-moment multi-temperature closure by its trace counterpart for heavy impurities at the trace limit, like for the tungsten case. Furthermore, one can also use the heuristic scheme for the thermal forces if the temperature ratio is close to unity. At this point, we mention a caveat that all treatment so far is only valid when radiative energy loss can be neglected, which may not be the case for heavy impurities even at low concentrations. Thus care must be taken to not use the any of the trace or full schemes in cases where radiation losses are significant. 

Now that we have tested the $21N$-moment multi-temperature Zhdanov closure for its convergence and performance against the trace approximation, we now proceed to relax some of the assumptions in order to illustrate a few specific cases of application.

\section{Relaxing some assumptions on the $21N$-moment multi-temperature closure}
\label{sec:relaxation}

The assumptions mentioned in Sec.\,\ref{sec:closure} generally are valid for the linear transport regime in the classical sense. However, it may so happen that there are systems where certain assumptions may need to be relaxed, since the ordering of various moments and their derivatives may become different due to unique physical conditions of that system. In this section, we consider two  special cases of interest and outline the assumptions which are relaxed. However we do not solve the equations here, as the solution method remains the same as the previous sections, and merely adds some extra terms to the linearized system of Eqs\,(\ref{eq:generalizedrank1}) and (\ref{eq:generalizedrank2}). And so, we proceed to describe the two special cases, the first being the extension of the closure to account for the terms present in the linearized Burnett approximation, from which we derive the second, i.e.\,the extension of the closure to account for certain comparable magnitude terms in the drift-approximation.

\subsection{$21N$-moment multi-temperature closure consistent to the linearized Burnett approximation}

In certain systems, it may so happen that the spatial gradients of the heat-flux $\mathbf{h}_\i$ and the stress tensor $\pi_\i$ are found in the first-order of the Knudsen number $K_n$, which corresponds to the linearized Burnett approximation\cite{chapman_mathematical_1952,zhdanov_transport_2002}. In such a case, the parallel $21N$-moment system gets modified as follows. Firstly, the equation for the heat-flux becomes
 \begin{equation}
  -\frac{5}{2}\frac{kT_\i}{m_\i}\nabla_\parallel.{\pi_{\i\parallel\parallel}}+ \frac{5}{2} \frac{k}{m_\i}n_\i kT_\i \nabla_\parallel T_\i=\sum_\j R^{11}_{\i\j\parallel}, \label{eq:21nhrelaxed}
\end{equation}
where the parallel divergence $(\nabla_\parallel.)=\mathbf{\hat{b}}(\mathbf{\hat{b}}.)(\nabla.)$, and the factor of $5/2$ since 
\begin{equation}
 \sigma_\i=\frac{m_\i}{2}\int c_\i^2\left(\mathbf{c}_\i\mathbf{c}_\i-\frac{1}{3}\bm{\delta}c_\i^2\right) f_\i d\mathbf{c}_\i-\frac{7}{2\gamma_\i}\pi_\i,
\end{equation}
and since we continue to neglect $\pi_{\i\parallel\parallel}.\nabla_\parallel T_\i$. The equation for the stress tensor similarly becomes
 \begin{multline}
 \frac{4}{5}\left(\frac{5}{2}n_\i kT_\i\left\{ \frac{\partial {w}_{\i r}}{\partial x_s}\right\}_{\parallel\parallel} +\left\{ \frac{\partial {h}_{\i r}}{\partial x_s}\right\}_{\parallel\parallel}\right)\\
 +2n_\i kT_\i\left\{\frac{\partial u_r}{\partial x_s}\right\}_{\parallel\parallel}=\sum_\j R_{\i\j\parallel\parallel}^{20},\nonumber
\end{multline}
which can be represented in terms of the usual heat flux $\mathbf{q}_\i$ as 
 \begin{equation}
 \frac{4}{5}\left\{ \frac{\partial {q}_{\i r}}{\partial x_s}\right\}_{\parallel\parallel}
 +2n_\i kT_\i\left\{\frac{\partial u_r}{\partial x_s}\right\}_{\parallel\parallel}=\sum_\j R_{\i\j\parallel\parallel}^{20},\label{eq:21npirelaxed}
\end{equation}
with the reduced balance equations for $n_\i b^{12}_\i=\mathbf{r}_\i$ and $n_\i b^{21}_\i=\sigma_\i$ remaining the same as earlier
\begin{equation}
 0=\sum_\j R^{12}_{\i\j \parallel},\ \text{and}\ 0=\sum_\j R^{21}_{\i\j \parallel\parallel}.\label{eq:21nextraequations}
\end{equation}
One can immediately see from Eqs.\,(\ref{eq:21nhrelaxed}) and Eq.\,(\ref{eq:21npirelaxed}) that they are no longer linear expressions amenable to direct solution through linear algebra techniques, and furthermore the expressions for rank-1 and rank-2 moments are coupled. Normally, they can be solved through the method of successive solutions, i.e. first determine a zeroth approximation ${h}_{\i\parallel}$ and $\pi_{\i \parallel\parallel}$ by ignoring the gradient and divergence terms in the $21N$-moment scheme, then substitute these solutions into Eqs.\,(\ref{eq:21nhrelaxed}) and Eq.\,(\ref{eq:21npirelaxed}) to obtain a first approximation, substitute the first approximation to obtain a second and so on until the solutions sufficiently converge. In principle, this requires one to perform the matrix inversion a sufficient number of times, and can be both algebraically tedious and computationally expensive. However, one can further assume that the higher order derivatives of the heat-flux and the stress-tensor are higher than order one in $K_n$, in case of which the first approximation will uncouple the equations, giving the solution expression for the heat-flux as 
 \begin{equation}
  -\frac{5}{2}\frac{kT_\i}{m_\i}\nabla_\parallel\cdot{
  (\eta_{\i\parallel\parallel0} \epsilon_{\parallel\parallel})}+ \frac{5}{2} \frac{k}{m_\i}n_\i kT_\i \nabla_\parallel T_\i=\sum_\j R^{11}_{\i\j\parallel},\label{eq:abc1}
\end{equation}
where $\eta_{\i\parallel\parallel0}$ represents the zeroth approximation for the partial viscosity $\eta_{\i\parallel\parallel}$ which we leave inside the divergence since it depends on the temperatures of the species. The parallel divergence term can be further expanded as 
\begin{equation}
 \nabla_\parallel.{(\eta_{\i\parallel\parallel0} \epsilon_{\parallel\parallel})}=\sum_\beta\frac{\partial \eta_{\i\parallel\parallel0}}{\partial T_\j} (\nabla_\parallel T_\j).\epsilon_{\parallel\parallel}+\eta_{\i\parallel\parallel0} \nabla_\parallel\cdot\epsilon_{\parallel\parallel}.
\end{equation}
The solution for the stress tensor becomes
 \begin{multline}
 \frac{4}{5}\sum_\j\left\{ \nabla\left( \lambda_{\j\i\parallel0}\nabla_\parallel T_\j+\alpha_{\j\i\parallel0}w_{\beta\parallel} \right)\right\}_{\parallel\parallel}\\
 +2n_\i kT_\i\epsilon_{\parallel\parallel}=\sum_\j R_{\i\j\parallel\parallel}^{20},\label{eq:abc2}
\end{multline}
where $\lambda_{\j\i\parallel0}$  is the zeroth solution for the partial parallel thermal conductivity and $\alpha_{\j\i\parallel0}$ can be termed the zeroth solution to the partial parallel thermomotive coefficient respectively (i.e. the coefficient of the heat-flux term depending on the diffusion velocities), which in turn depend on the temperatures of the species and can be expanded similarly as above in terms of gradients of temperature for the definition of the partial thermoelectric coefficients). These equations can now be solved individually rank-wise to obtain the first approximations to the partial viscosities $\eta_{\i\parallel\parallel1}$ and partial thermal conductivities $\lambda_{\j\i\parallel1}$ (and partial thermoelectric coefficients similarly). The system of equations (\ref{eq:abc1}) and (\ref{eq:abc2}) may be considered the multi-temperature generalization of Eq. (5.2.21) of Ref.\,\onlinecite{zhdanov_transport_2002}, and generally finds use in extended nonequilibrium thermodynamics, where the moment method is used for relating generalized thermodynamic forces to the thermodynamic fluxes\cite{jou_extended_1988,jou_extended_1999,zhdanov_method_1998,zhdanov_kinetic_2002}. Together with Eq.\,(\ref{eq:21nextraequations}), they may be considered the extended $21N$-moment closure. On the basis of this, we now proceed to illustrate the closure in the drift approximation. 

\subsection{$21N$-moment multi-temperature scheme in the drift-approximation}

Following neoclassical theory, we can separate out flows as $\mathbf{u}_\i=\mathbf{u}_{\i0}+\mathbf{u}_{\i1}$, where the zeroth order flow $\mathbf{u}_{\i0}$ consists of the individual species flow and $\mathbf{E}\times\mathbf{B}$ flow. The commmon flow $\mathbf{w}_\i$ now would be calculated by the density weighted average of $\mathbf{u}_{\i0}$. In the drift approximation of Hinton and Hazeltine$^{35,36}$, especially for tokamak regions with strong pressure gradients, it may so happen that the total first-order flow velocity $\mathbf{u}_{\i1}= \mathbf{u}_{\i1,\perp}  + \mathbf{u}_{\i1,\parallel}$, where $\mathbf{u}_{\i1,\perp}=\mathbf{B}\times\nabla p_\i/n_iZ_ieB^2$ is the diamagnetic flow velocity and $\mathbf{u}_{\i1,\parallel}$ is its parallel return flow, is comparable to the main particle flow and the $\mathbf{E}\times\mathbf{B}$ flows. Accounting for this flow is important, since it often is the main contributor to the augmented friction force that leads to impurity accumulation in steady state with absence of turbulence$^{37,38}$. Furthermore, this diamagnetic flow is usually accompanied by the corresponding first-order neoclassical heat-flux$^{39}$ $\mathbf{q}_{\i1}= \mathbf{q}_{\i1,\perp}  + \mathbf{q}_{\i1,\parallel}$, where the diamagnetic heat-flux is $\mathbf{q}_{\i1,\perp} =(5/2)p_\i\mathbf{B}\times\nabla p_\i/n_iZ_ieB^2$ and $\mathbf{q}_{\i1,\parallel}$ the corresponding neoclassical parallel heat-flux of the first-order. The gradients of this first-order heat flux may indeed be significantly larger than the gradients of particle fluid flow velocity and the $\mathbf{E}\times\mathbf{B}$ flow velocity. And indeed any further gradient of the gradient of the diamagnetic heat-flux would be an order higher, and hence can be neglected. Furthermore, the Knudsen number ordering scheme of the Zhdanov closure, other than for the first-order heat-flux and flow, is compatible with the drift ordering. Thus, in such a situation, one only needs to incorporate the diamagnetic heat-flux in Eqs.\,(56) and (58). In case of this, the expressions can be modified as follows to find
 \begin{multline}
  -\frac{5}{2}\frac{kT_\i}{m_\i}\nabla_\parallel\cdot{
  (\pi_{\i0}+\pi_{\i1})}+ \frac{5}{2} \frac{k}{m_\i}n_\i kT_\i \nabla_\parallel T_\i\\
  =\sum_\j (R^{11}_{\i\j0,\parallel}+M^{110}_{\i\j}{u}_{\j1,\parallel}+M^{111}_{\i\j}{h}_{\j1,\parallel}+M^{112}_{\i\j}{r}_{\j1,\parallel}), \label{eq:21nhdia} \tag{64}
\end{multline}
and
 \begin{multline}
 \frac{4}{5}\left(\left\{\nabla \mathbf{q}_{\i0}\right\}_{\parallel\parallel}+\left\{\nabla \mathbf{q}_{\i1}\right\}_{\parallel\parallel}\right)\\
 +2n_\i kT_\i(\epsilon_{\parallel\parallel}+\epsilon_{\i1,\parallel\parallel})=\sum_\j (R_{\i\j\parallel\parallel}^{20}+M^{200}_{\i\j}{\pi}_{\j1,\parallel\parallel}),\label{eq:21npidia}\tag{65}
\end{multline}
where $\epsilon_{\i1,\parallel\parallel}=\left\{\nabla\mathbf{u}_{\i1}\right\}_{\parallel\parallel}$. The equations for $\mathbf{r}_\i$ is similarly modified,  
\begin{equation}
 0=\sum_\j (R^{12}_{\i\j \parallel}+M^{120}_{\i\j}{u}_{\j1,\parallel}+M^{121}_{\i\j}{h}_{\j1,\parallel}+M^{122}_{\i\j}{r}_{\j1,\parallel}),\tag{67}
\end{equation}
but the equation for $\sigma_\i$ contains a new term, as follows
\begin{multline}
\frac{4}{5}\left(\left\{\nabla \mathbf{s}_{\i0}\right\}_{\parallel\parallel}+\left\{\nabla \mathbf{s}_{\i1}\right\}_{\parallel\parallel}\right)\\
+\frac{n_\i k}{m_\i}\frac{14}{5}\left\{(\mathbf{q}_{\i0}  + \mathbf{q}_{\i1})\nabla T_\i \right\}_{\parallel\parallel}
=\sum_\j (R^{21}_{\i\j \parallel\parallel}+M^{210}_{\i\j}{\pi}_{\j1,\parallel\parallel}),\tag{68}\label{eq:68}
\end{multline}
where $\mathbf{s}_\i$ is a moment given by $\mathbf{s}_\i=\int (m_\i c_\i^4/4)\mathbf{c}_\i f_\i d\mathbf{c}_\i$. A simplification of Eq.\,(\ref{eq:68}) can be performed by Grad's closure on the moment $\mathbf{s}_{\i}$ therfore setting $\mathbf{s}_{\i}=(7/2\gamma_\i)\mathbf{h}_{\i}$. The first-order parallel quantities can be substituted for the ones in the relevant neoclassical regime.
A possible method of solution for such a system can be found in the solution of a similar system in Ref.\,13, in which, the equations are simplified so as to make the rank-1 equations solvable first by neglecting the stress-tensor divergence term, and then rank-2 equations are solved including the heat-flux term and by Grad's closure on $\mathbf{s}_{\i}$.\footnote{We would like to thank S.O.\,Makarov (Max-Planck Institut fur Plasmaphysik, Germany) for pointing out some errors and suggesting some clarifications from the beginning of this subsection until this point. For the differences, please compare with the earlier arXiv version.} It is also worth noting that including $\mathbf{q}_{\i1}$ in such a manner, either in the momentum equation or the closure or both, can lead to inconsistencies\cite{gath_consistency_2019} such as summabilitity issues\cite{poulsen_turbulent_2020}. That is, if a species is split into two continuous portions all else being equal, the two split portions may not evolve together as the whole unsplit species would. However, it becomes more and more consistent as the pressure profiles of various species approach each other.

We can see that this form of the closure still remains quasi-linear (i.e.,at least linear in the derivatives), and it can be still solved through linear algebra techniques since the diamagnetic flow and its associated diamagnetic heat flux are fully determined (on knowing the pressure and density profiles and the magnetic field).

\section{Summary of conclusions and outlook}
\label{sec:summary}

We first begin with the Sonine-Hermite polynomial moment method of Grad\cite{grad_asymptotic_1963} and Zhdanov\cite{alievskii_1963_transport,zhdanov_transport_2002}, and using the methods described wherein, we re-derive and verify the most general moment-averaged balance equation presented in Appendix A of Ref.\,\onlinecite{zhdanov_transport_2002}. We also explain certain subtleties of notation, leading to additional terms not present in Ref.\,\onlinecite{zhdanov_transport_2002}. Using this general balance equation for the basis of linearization, we outline the assumptions behind the linearization of this system of balance equations, with detailed assumptions behind the order of terms, on the basis of which terms are retained and discarded. We note here generally that the plasmadynamical moments are considered of a zeroth order in Knudsen number $K_n$, and the time derivative as well as the gradients of the plasmadynamical moments, and the higher-order moments are generally considered to be of first order in $K_n$. The higher-order time derivatives and gradients of the plasmadynamical moments, and the first and higher-order time derivatives and the gradients of higher-order moments, are all considered to be of or smaller than the order $(K_n)^2$. 

On the basis of these approximations, we illustrate a general linearization scheme (following, but not exactly the same as, Secs.\,4.6 and 6.3 of Ref.\,\onlinecite{zhdanov_transport_2002}), which leads to a hierarchy of higher-order parallel moment equations such that the only equations with a non-vanishing LHS are the ones for the parallel heat-flux and the longitudinal stress tensor. The rest of the higher-order moment equations only contain the collisional RHS terms. This leads to the system of equations rendered solvable by linear algebra techniques. The solution expresses the higher-order parallel vectorial moments in terms of the flow velocities and temperature gradients, and express the higher-order longitudinal tensorial moments in terms of the rate-of-strain tensor, which resemble the classical transport solutions in their usual form. In particular, the coefficient of the parallel temperature gradients in the parallel heat-fluxes are the partial parallel thermal conductivities, and the coefficient of the longitudinal rate-of-strain tensor in the longitudinal stress tensor is the partial longitudinal viscosity. Furthermore, we illustrated the calculation of the generalized friction and thermal forces using the expressions for the higher-order moments found in this manner, which is what is key for implementation in code packages which simulate SOL/edge plasmas. Thus, having illustrated the general Zhdanov closure, we proceeded to study its convergence with respect to addition of a number of extra moments. To quantify this a little better for different impurities found in fusion, in addition to deuterium-tritium, we chose Carbon and Argon, mid-weight impurities at significant fraction of main ion density, and Tungsten, a heavy impurity at trace levels (Table I).

We choose seven rank-1 moments and eight rank-2 moments in order to test the convergence of the closure scheme. On calculating the total viscosity and the total thermal conductivities, the first observation we made was that the most significant addition to the viscosity and thermal conductivities seems to be indeed on going from the $13N$-moment scheme to the $21N$-moment scheme. The addition of further moments causes much smaller changes compared to the $21N$-moment scheme. However, addition of more moments still leads to convergence of the scheme, particularly in representing the transport coefficients precisely to more decimal places. The total viscosity, in particular, seems to converge at a much faster rate than the total thermal conductivities. At this point, we chose an arbitrary two decimal places of precision to provide recommendations. We find again that the maximum differences for any transport coefficient at any number of moments are found for the mid-weight impurities calculated with the single-temperature coefficients. 
This is especially pronounced for the thermal conductivities, on the basis of which we recommend using multi-temperature $21N$-moment scheme for the mid-weight impurities to achieve the two decimal places precision. We furthermore observe that the differences between the single-temperature transport coefficients and the multi-temperature transport coefficients seem to still remain significant on addition of more moments, between $10-80\%$, with the largest differences for the mid-weight impurities. 
Furthermore, we also calculate the augmentation of the friction and thermal forces caused by the additions of more moments. As with the convergence of transport coefficients, we find that the friction forces are more-or-less converged to two decimal places in the $21N$-approximation, and the thermal forces to one decimal place or so. For the mid-weight impurities, particularly argon, the convergence for the thermal force was much slower. Furthermore, significant difference in the force values from the single-temperature and multi-temperature coefficients still persisted on addition of more moments. 
On the basis of this, we generally recommend using multi-temperature coefficients where feasible, in addition to the earlier recommendations. 

Having found the $21N$-moment scheme suitable for most cases, we proceeded to study the performance of the trace approximation against it. The trace approximation generally assumes the impurity particles to be present at trace values with respect to the background plasma density, and hence the trace approximation entails neglecting the impurity self-collisions and the back reaction of the impurities on the main plasma. On doing so, and on comparing the transport coefficients and forces, we found generally that the trace approximation tends to overestimate the transport coefficients and thermal forces by at least a factor of two for the mid-weight and low-weight impurities at significant densities. For the trace tungsten case, we found that the transport coefficients and the forces practically overlapped, justifying the trace limit. Generally, we also found that there persisted significant differences between the single-temperature and multi-temperature cases as well (at least 40
\%). On the basis of these, we can recommend using the multi-temperature trace approximation on the $21N$-moment Zhdanov closure for heavy impurities at trace levels in order to  speed up computations in fluid codes.  

At last, we also described methods by which some of the assumptions may be relaxed, particularly two cases, the first in which the closure needs to respect the linearized Burnett approximation, and following which, the closure in the drift approximation. We did not compute the transport coefficients for these relaxed cases, as essentially the matricial method remains the same, but just adds more relatively constant terms.

In the scope of this article, we studied closures in the multi-species linear transport regime to obtain parallel transport coefficients, but we did not study the perpendicular ones. There are some analytical issues when a magnetic field is explicitly introduced in the multi-species system of equations in the linear transport regime, such as the increase in the degrees of freedom on which the transport parameters depend, as opposed to that of the ion-electron case where the degree of freedom essentially boils down to one variable $\beta_\i=\Omega_\i\tau_{\i\i}$\cite{balescu_transport_1988,ji_closure_2013}. The extension of the multi-species closure scheme to the fully magnetized case while addressing such issues, in order to study the perpendicular transport coefficients, is a subject of our future work. Though in the perpendicular direction, transport is often dominated by turbulence, they may still be useful to provide a more precise neoclassical description, which in recent literature has been found to account for impurity transport very well in absence of turbulence and MHD modes\cite{angioni_impurity_2021}. We also plan to consider the effects of parallel electric field on the closure and study the related thermoelectric effects. Furthermore, the scope of the study was committed to a fully analytical examination of the transport coefficients, and no concrete effects on numerical modelling were mentioned other than the suggested ranges of validity. The effects of the multi-species closure scheme with multi-temperature coefficients on numerical SOL/edge fluid packages, Soledge3x-EIRENE in our case, is a subject we will explore.

\section*{Acknowledgments}

The projects leading to this publication have received funding from Excellence Initiative of Aix-Marseille Universit\'e - A*MIDEX, a French Investissement d'Avenir Programme, project TOP \& AMX-19-IET-013. 

\section*{Data Availibility}

The data in the article can be made available by the authors on reasonable request.

\bibliography{bibliography}{}
\bibliographystyle{vancouver}

\appendix

\section{Polynomial Identities}
\label{sec:polynomialidentities}

We we collect together some properties for the polynomials provided by Zhdanov\cite{zhdanov_transport_2002} and Weinert et al\cite{weinert_spherical_1980}, rewritten in our notation.

\subsection{Identities for Sonine polynomials $S^n_m$}

The Sonine polynomials $S^n_m(x)$ follow the recurrence relations
\begin{align}
 S^n_{m+1/2}\left(x\right) &= S^n_{m+3/2}\left(x\right)-S^{n-1}_{m+3/2}\left(x\right)\\ 
 x S^n_{m+1/2}\left(x\right) &= \left(n+m+\frac{1}{2}\right)S^n_{m-1/2}\left(x\right) -(n+1)S^{n+1}_{m-1/2}\left(x\right).
\end{align} 
for any scalar $x$, The derivative with respect to $x$ is given by
\begin{equation}
 \frac{d}{d x}S^n_{m+1/2}\left(x\right)=-S^{n-1}_{m+3/2}\left(x\right).
\end{equation}
The gradient w.r.t.\,$\mathbf{c}_\i$ is given by
\begin{equation}
 \frac{d}{d \mathbf{c}_\alpha}S^n_{m+1/2}\left(\frac{\gamma_\alpha}{2}\mathbf{c}^2_\alpha\right)=-{\gamma_\alpha \mathbf{c}_\alpha}S^{n-1}_{m+3/2}\left(\frac{\gamma_\alpha}{2}\mathbf{c}^2_\alpha\right).
\end{equation}
And finally, the derivative with respect to $\gamma_\i$ is given by
 \begin{equation}
 \frac{d}{d \gamma_\alpha}S^n_{m+1/2}\left(\frac{\gamma_\alpha}{2}\mathbf{c}^2_\alpha\right)=-\frac{c_\alpha^2}{2}S^{n-1}_{m+3/2}\left(\frac{\gamma_\alpha}{2}\mathbf{c}^2_\alpha\right).
\end{equation}

\subsection{Identities for the Irreducible tensorial monomial $P^{(m)}$}

For the irreducible monomial $P^{(m)}(\bm{\xi})$, the following relations, all in the symmetrization notation, all but the last from Ref.\,\onlinecite{zhdanov_transport_2002}
\begin{enumerate}
\item  The Rodrigues form of the irreducible tensorial monomial $P^{(m)}(\bm{\xi})$ is given by
\begin{equation}
   P^{(m)}(\bm{\xi})=\frac{1}{(2m-1)!!}\xi^{2m+1}(-\nabla_{\bm{\xi}}^m)\left(\frac{1}{\xi}\right),
 \end{equation}
where $n!!$ indicates the double factorial of $n$. Sometimes, the symbol for the spherical tensor $Y_m(\bm{\xi})$ is used in place of $P^{(m)}(\bm{\xi})$, as they are both the same. (See Appendix \ref{sec:symmetrizationnote}) for why the spherical tensor is equivalent to the irreducible monomial.)
 \item Contraction with $\xi_r$, i.e.\,1-fold inner product with $\bm{\xi}$, $\bm{\xi}\cdot P^{(m)}_{a_1 a_2\ldots a_r\ldots a_m}$,
 \begin{equation}
  \xi_rP^{(m)}_r=\frac{m\xi^2}{2m-1}P^{(m-1)}.
 \end{equation}
 This is essentially the same as the contraction identity of the spherical tensor $Y_m\equiv P^{(m)}$ in Ref.\,\onlinecite{weinert_spherical_1980}. 
 \item Gradient $\nabla P^{(m)}$
 \begin{multline}
  \frac{\partial P^{(m)}}{\partial \xi_r}=\left\{P^{(m-1)}\bm{\delta} \right\}_r\\
  =\bm{\delta}_r P^{(m-1)}-\frac{2}{2m-1}\bm{\delta}P_r^{(m-1)},\label{eq:pmgradient}
 \end{multline}
 where $\bm{\delta}_r$ refers to symmetrization without the participation of the $r^{th}$ index. E.g.\, $\bm{\xi}\bm{\delta}_r=\xi_s\delta_{rt}+\xi_t\delta_{rs}$, and $\bm{\delta}_r\bm{\delta}_s=\delta_{ru}\delta_{st}+\delta_{rt}\delta_{su}$.
 
 Additionally, one can also define derivative with respect to $\gamma$, if $\bm{\xi}=\gamma^{1/2}\mathbf{c}$,
 \begin{equation}
  \frac{\partial P^{(m)}}{\partial \gamma}=\frac{m}{2\gamma}P^{(m)}.
 \end{equation}

 \item Outer product with $\bm{\xi}$, $\bm{\xi} P^{(m)}$,
 \begin{equation}
  {\xi}_r P^{(m)}=P^{(m+1)}+\frac{\xi^2}{2m+1}\frac{\partial P^{(m)}}{\partial \xi_r},
 \end{equation}
 or
 \begin{equation}
  {\xi}_r P^{(m)}=P^{(m+1)}+\frac{\xi^2}{2m+1}\left\{P^{(m-1)}\bm{\delta} \right\}_r.
 \end{equation}
 
 This identity can also be verified as in Refs.\,\onlinecite{weinert_spherical_1980,ji_exact_2006,jorge_linear_2019}, through using the Rodrigues form of the spherical tensor directly.

 \item Tensorial double gradient $\bm{\nabla}^2 P^{(m)}$
 \begin{multline}
  \frac{\partial^2}{\partial \xi_r \partial \xi_s}P^{(m)}=\left\{P^{(m-2)}\bm{\delta}\bm{\delta} \right\}_{rs}=\frac{2m+1}{2m-1}\delta_r\delta_s P^{(m-2)}\\
  -\frac{2}{2m-1}\bm{\delta\delta} P^{(m-2)}+\frac{2(2m+1)}{(2m-1)(2m-3)}\bm{\delta\delta} P^{(m-2)}_{rs}.
 \end{multline}
 Note the difference from the identity provided in the appendix of Ref.\,\onlinecite{zhdanov_transport_2002}.
 \item Outer product of the gradient with $\bm{\xi}$, $\bm{\xi}\nabla P^{(m)}$,
 \begin{multline}
  \xi_s  \frac{\partial P^{(m)}}{\partial \xi_r}=\left\{P_s^{(m)}\bm{\delta} \right\}_r+\frac{\xi^2}{2m-1}\left\{P^{(m-2)}\bm{\delta\delta} \right\}_{rs}.
 \end{multline}
 \item Outer product with $\bm{\xi}^2$
 \begin{multline}
  \xi_r \xi_s P^{(m)}= P^{(m+2)} +\frac{\xi^2}{2m+1} \{\bm{\delta}_rP^{(m)}\}_s+\frac{\xi^2}{2m+1}\{\bm{\delta}_sP^{(m)}\}_r\\
  +\frac{\xi^4}{4m^2-1}\left\{P^{(m-2)}\bm{\delta\delta} \right\}_{rs}.
 \end{multline}

\end{enumerate}
 For all the above identities,  any $r^{th}$ or $(rs)^{th}$ component of a tensor mentioned as $P^{(m)}_r$ or $P^{(m)}_{rs}$ refers to the additional ranks the tensor has, i.e. the rank-0 and rank-1 quantities do not possess an additional $r^{th}$ or $(rs)^{th}$ component, and rank-2 quantities do not possess an additional $(rs)^{th}$ component. In case such identities call for terms which have $P^{(0\ \mathrm{or}\ 1)}_{r\ \mathrm{or}\ rs}$ or $P^{(2)}_{rs}$, such terms can be set to zero.                                                                                                                                                                                                                                                                                                                                                                                                                                                                                                                                                                                                                                                                                                                                                                                                                                                                                                                                                                                                                                                                                                                                                                                                                                                                                                                                                                                                                                                                                                                                                                                                                                                                                                                                                                                                                                                                                                                                                                                                                                                                                                                                                                                                                                                                                                                                                                                                                                                                                                                                                                                                               

\subsection{Identities for the Sonine-Hermite polynomial $G^{mn}$}

The following relations for the Sonine-Hermite polynomials as given in Ref.\,\onlinecite{zhdanov_transport_2002}, re-written in consistent terms, with derivation methods wherever needed
\begin{enumerate}
 \item The derivative with respect to $\gamma_\i$ becomes
 \begin{equation}
  \frac{\partial G^{mn}_\i}{\partial \gamma_\i}=\frac{1}{2\gamma_\i^2}n(2m+2n+1)G^{m,n-1}_\i \label{eq:gmnderivativegamma}
 \end{equation}
  \item Outer product with $\bm{c}_\i$
 \begin{multline}
  c_{\i s}G^{mn}_\i=G^{m+1,n}_\i +\frac{n}{\gamma_\i} G^{m+1,n-1}_\i\\
  +\frac{2}{2m+1}\left[\frac{2m+2n+1}{2\gamma_\i}\{G^{m-1,n}_\i\bm{\delta}\}_s +\{G^{m-1,n+1}_\i\bm{\delta}\}_s     \right]\label{eq:gmnouterproductc}
 \end{multline}
 \item Outer product with $\mathbf{c}_\i$ of the derivative w.r.t.\,$\gamma_\i$
 \begin{multline}
  c_{\i s}\frac{\partial G^{mn}_\i}{\partial \gamma_\i}=\frac{n(2m+2n+1)}{2\gamma_\alpha^2}\times\\
  \times\left[G^{m+1,n-1}_\i+\frac{n-1}{\gamma_\i}G^{m+1,n-2}_\i\right.\\
  \left.+\frac{2}{2m+1}\left(\frac{2m+2n-1}{2\gamma_\i}\{G^{m-1,n-1}_\i\bm{\delta}\}_s + \{G^{m-1,n}_\i\bm{\delta}\}_s\right) \right]
 \end{multline}
 This follows straightforwardly from the above two identities Eqs.\,(\ref{eq:gmnderivativegamma}) and (\ref{eq:gmnouterproductc}).
 \item Gradient with respect to $\bm{c}_\i$ 
 \begin{equation}
  \frac{\partial G^{mn}_\i}{\partial c_{\i s}} = nG^{m+1,n-1}_\i +\frac{2m+2n+1}{2m+1}\{G^{m-1,n}_\i\bm{\delta}\}_s
 \end{equation}

 \item Outer product of gradient w.r.t.\,$\bm{c}$ with $\bm{c}$
 \begin{multline}
   c_{\i s}\frac{\partial G^{mn}_\i}{\partial c_{\i r}}=nG^{m+2,n-1}_\i +\frac{n(n-1)}{\gamma_\i}G^{m+2,n-2}_\i+\frac{2n}{2m+3}\bm{\delta}G^{mn}_\i\\
   +\frac{n(2m+2n+1)}{(2m+3)\gamma_\i}\bm{\delta}G^{m,n-1}_\i+\{G^{mn}_{\i s}\bm{\delta}\}_r-\frac{2n}{2m-1}\bm{\delta}G^{mn}_{\i rs}\\
   -\frac{n(2m+2n+1)}{(2m-1)\gamma_\i}\bm{\delta}G^{m,n-1}_{\i rs}  +\frac{2m+2n+1}{4m^2-1}\times\\
   \times\left[2\{G^{m-2,n+1}_\i\bm{\delta\delta}\}_{rs}+\frac{2m+2n-1}{\gamma_\i}\{G^{m-2,n}_\i\bm{\delta\delta}\}_{rs} \right].
 \end{multline}
 Notice the difference of the coefficient of $\bm{\delta}G^{mn}_{\i rs}$, and $\{G^{mn}_{\i s}\bm{\delta}\}_r$ instead of $G^{mn}_{\i s}\bm{\delta}_r$ from the expression in the appendix of Ref.\,\onlinecite{zhdanov_transport_2002}. Notice the difference of a minus sign in $2m+2n-1$ in the coefficient of $\{G^{m-2,n}_\i\bm{\delta\delta}\}$, note the specification of the $rs^{th}$ components in the $\bm{\delta}G^{m,n-1}_{\i rs}$ term.
 \item Double outer product with $\bm{c}$
 \begin{multline}
  c_{\i s}c_{\i r}G^{mn}_\i=G^{m+2,n}_\i+\frac{2n}{\gamma_\i}G^{m+2,n-1}_\i\\
  +\frac{c_\i^2}{2m+1}\{\bm{\delta}_rG^{mn}_\i\}_s+\frac{c_\i^2}{2m+1}\{\bm{\delta}_sG^{mn}_{\i}\}_r\\
  +\frac{1}{4m^2-1}\left[\frac{(2m+2n+1)(2m+2n-1)}{\gamma_\i^2}\{G^{m-2,n}\bm{\delta\delta}\}_{rs} \right.\\
  \left.+\frac{4(2m+2n+1)}{\gamma_\i}\{G^{m-2,n+1}\bm{\delta\delta}\}_{rs}+4\{G^{m-2,n+2}\bm{\delta\delta}\}_{rs}\right]
 \end{multline}

\end{enumerate}

\section{A note on symmetrization notation}
\label{sec:symmetrizationnote}

In the existing literature on symmetric and irreducible tensorial polynomials, there exist multiple notations, which often differ from one another by rational factors. This often makes the task of verifying expressions written in different notations difficult. Here we provide a short description of the three main notation systems encountered. The first form, notated without any special symbols, as used by Grad\cite{grad_asymptotic_1963,grad_note_1949} and Zhdanov\cite{zhdanov_transport_2002} write terms directly. For example, in this notation, $\bm{\xi}\bm{\delta}$ is already symmetrized. For example
\begin{align}
\{\bm{\delta}\}_{ij}&=\delta_{ij},\ \{\bm{\delta}^2\}_{ijkl}=\delta_{ij}\delta_{kl}+\delta_{ik}\delta_{jl}+\delta_{il}\delta_{jk}\\
 \{\bm{\xi}\bm{\delta}\}_{ijk}&=\xi_i\delta_{jk}+\xi_j\delta_{ik}+\xi_k\delta_{il} 
\end{align}
and so on. Some care has to be exercised with the notation, as in the example provided by Grad himself,
\begin{equation}
 (\bm{\xi^2}+\bm{\delta})\bm{\delta}=\bm{\xi^2}\bm{\delta}+2\bm{\delta^2},
\end{equation}
because $\bm{\delta}\bm{\delta}$ repeats symmetrized quantities in $\bm{\delta^2}$ twice. In such a notation, performing an outer multiplication of symmetrized irreducible rank-$m$ tensor with $\bm{\xi}$ will need to be accompanied by a factor of $1/(m+1)$ as such a multiplication would repeat the terms in the new rank-$(m+1)$ tensor $(m+1)$ times.  

The second notation involves the use of parentheses $(...)$ in the indices to indicate symmetrization. For example, we have the following
\begin{align}
\delta_{(ij)}&=\delta_{ij},\ \delta_{(ij}\delta_{kl)}=\frac{1}{3}(\delta_{ij}\delta_{kl}+\delta_{ik}\delta_{jl}+\delta_{il}\delta_{jk})\\
 \xi_{(i}\delta_{jk)}&=\frac{1}{3}(\xi_i\delta_{jk}+\xi_j\delta_{ik}+\xi_k\delta_{il}) 
\end{align}
For a general $\xi_{(i_1}...\xi_{i_n}\delta_{j_1j_2}...\delta_{j_{2m-1}j_{2m})}$, the relation between the two notations is given by
\begin{equation}
 \bm{\xi^n}\bm{\delta^m}=\frac{(n+2m)!}{(2!)^m m!n!}\xi_{(i_1}...\xi_{i_n}\delta_{j_1j_2}...\delta_{j_{2m-1}j_{2m})}.
\end{equation}
Proof of this relationship can be derived from simple combinatorics of indices. With this relation, one can verify that the Grad-Ikenberry polynomials\cite{ikenberry_system_1955,ikenberry_system_1961,ikenberry_representation_1962}, the spherical tensor\cite{weinert_spherical_1980,johnston_general_1966} and Grad's symmetric irreducible tensorial polynomials $P^{(m)}$ are one and the same, a fact often not clearly mentioned in previous literature. 

The third notation encountered is the use of external curly brackets $\{...\}$ on the terms being multiplied to indicate symmetrization. For example $\{\bm{\xi^n}\bm{\delta^m}\}$ indicates the symmetrization of the tensorial outer product of $\bm{\xi^n}$ with $\bm{\delta^m}$. This is essentially the same as the second notation $\xi_{(i_1}...\xi_{i_n}\delta_{j_1j_2}...\delta_{j_{2m-1}j_{2m})}$. It is sometimes found in quantum mechanics and continuum mechanics literature, but often just for notating simple symmetrization, e.g.\,of the sort when there are two tensors $\mathbf{A}$ and $\mathbf{B}$ of the rank-1, then the symmetric dyad $\{\mathbf{AB}\}=(\mathbf{AB}+\mathbf{BA})/2$. However, this is not so easy to understand for more complicated symmetrizations, where an index based notation is better. 

In this article, we largely only use the first notation, and specify whenever it is used. The second notation and third notations are never used. However, a simple transformation can be performed in case required, using the abovementioned relationships between the different notations. However, we sparingly use curly brackets to denote symmetric, irreducible dyads, mentioning the definition when it is used. It is not to be confused with the third symmetrization notation.

\section{Balance equations}
\label{sec:balanceequations}

One can derive balance equations for any given moment by substituting the appropriate values of $(m,n)$ in Eq.\,(\ref{eq:generallhs}). In this appendix, we derive a few specific cases for the purposes of verification of the LHS with known results, deriving general balance equations for scalar, vector and tensor moments, and finally for deriving the full balance equations for specific moments missing from previous literature.

\subsection{$13N$-moment system of balance equations}
\label{subsec:13n}

The balance equation for the density $n_\i b^{00}_\i=\rho_\i$ is given by
 \begin{equation}
 \frac{d\rho_\alpha}{dt}+\rho_\alpha\nabla\cdot\mathbf{u}
 +\frac{\partial}{\partial x_r}(\rho_\i w_{\i r})=\sum_\j R_{\i\j}^{00}. \label{eq:13n1}
\end{equation}
Next, the balance equation for diffusion momentum $n_\i b^{10}_\i=\rho_\i\mathbf{w}_\i$
 \begin{multline}
 \frac{d}{dt}(\rho_\alpha \mathbf{w}_\i)+\rho_\alpha \mathbf{w}_\i\nabla\cdot\mathbf{u}
 +\frac{\partial}{\partial x_r}(\pi_{\i r}+n_\i kT_\i)\\
 -\rho_\i\left(\frac{\mathbf{X}_{\i}}{m_\i}+\frac{Z_\i e\mathbf{E}}{m_\i}+\mathbf{u}\times\bm{\omega}_\i-\frac{d \mathbf{u}}{dt}\right) \\
 +\rho_\i\left(w_{\i s}\frac{\partial u_r}{\partial x_s}-\mathbf{w}_\i\times\bm{\omega}_\i  \right)      =\sum_\j R^{10}_{\i\j} \label{eq:13n2}
\end{multline}
Similarly, the balance equation for $n_\i b^{01}_\i=0$, which effectively results in a balance equation for the temperature $T_\i$ is given by, 
 \begin{multline}
 \frac{3}{2}k n_\i\frac{d T_\i }{dt}+\frac{3}{2}k n_\i w_{\i r}\frac{\partial T_\i}{\partial x_r}+\frac{\partial}{\partial x_r}(h_{\i r} +n_\i kT_\i w_{\i r})\\
 -\left(\frac{X_{\i l}}{m_\i}+\frac{Z_\i eE_{\i l}}{m_\i}+\{\mathbf{u}\times\bm{\omega}_\i\}_l-\frac{d u_l}{dt}\right)\rho_\i w_{\i l}\\
 +\pi_{\i rs}\left\{\frac{\partial u_r}{\partial x_s} \right\} +n_\i kT_\i \nabla.\mathbf{u}
   = \sum_\j R^{01}_{\i\j}, \label{eq:13n3}
\end{multline}
from the fact that $\epsilon_{rst}\delta_{rs}$ vanishes, and that $\epsilon_{rst} \pi_{\i rs}\omega_{\i t}$ vanishes because $\pi_{\i rs}$ is symmetric. The curly brackets $\{AB\}$ represent symmetric reduction of a dyad $AB$, e.g.
\begin{equation}
 \{A_rB_s\}=\frac{1}{2}(A_rB_s+B_rA_s)-\frac{1}{3}\delta_{rs}A_lB_l.\nonumber
 \end{equation}
Notice that the evolution of temperature depends on the magnetic field only through the drift produced by the common flow. So far, the plasmadynamical equations take their usual form. 

The balance equation for the heat flux $n_\i b^{10}_\i=\mathbf{h}_\i$ is calculated from
 \begin{multline}
 \frac{d\mathbf{h}_\i}{dt}+\mathbf{h}_\i\nabla\cdot\mathbf{u}
 +\frac{\partial}{\partial x_r}\left[\sigma_{\i r} +\frac{1}{\gamma_\i}\pi_{\i r}+\frac{2}{3}\left(\{n_\i b^{02}_\i\bm{\delta}\}_r     \right) \right]\\
 -\left(\frac{X_{\i l}}{m_\i}+\frac{Z_\i eE_{\i l}}{m_\i}+\{\mathbf{u}\times\bm{\omega}_\i\}_l-\frac{d u_l}{dt}\right)\pi_{\i l}\\
 -\left(\epsilon_{rst}\omega_{\i t}+\frac{\partial u_r}{\partial x_s} \right)\left[ n_\i b^{30}_{\i rs} +\frac{2}{5}\bm{\delta}\mathbf{h}_\i+n_\i kT_\i\bm{\delta}\mathbf{w}_\i+\{\bm{\delta} h_{\i s}\}_r\right]\\
   +\frac{5n_\i k}{2}\mathbf{w}_\i\frac{d T_\i }{dt}\\
   +\frac{5 k}{2m_\i}\left[\pi_{\i r}+n_\i kT_\i\{\bm{\delta}\}_r \right]\frac{\partial T_\i}{\partial x_r}=\sum_\j R^{11}_{\i\j}, \nonumber
\end{multline}
where $s,t,l$ are repeated indices summed over, and the free index $r$ is used to show the remaining vectorial term from the double dot products. Now defining the third scalar moment $n_\i b_\i^{02}=\theta_\i$, the first rank-3 moment $n_\i b_\i^{30}=\text{\CYRZH}_\i$, and performing the expansions of $\bm{\delta}\mathbf{h}_\i$ and $\bm{\delta}\mathbf{w}_\i$ and their double dot products with $\partial u_r/\partial x_s$, we get 
 \begin{multline}
 \frac{d\mathbf{h}_\i}{dt}
 +\frac{\partial}{\partial x_r}\left(\sigma_{\i r} +\frac{1}{\gamma_\i}\pi_{\i r}\right)+\frac{2}{3}\nabla\theta_\i     \\
 -\pi_{\i l}\left(\frac{X_{\i l}}{m_\i}+\frac{Z_\i eE_{\i l}}{m_\i}+\{\mathbf{u}\times\bm{\omega}_\i\}_l-\frac{d u_l}{dt}\right)\\
 -\mathbf{h}_\i\times\bm{\omega}_\i+\text{\CYRZH}_{\i rs}\left\{\frac{\partial u_r}{\partial x_s} \right\}\\
 +\frac{7}{5}h_{\i s}\frac{\partial u_r}{\partial x_s}+\frac{2}{5}h_{\i t}\frac{\partial u_t}{\partial x_r}+\frac{7}{5}\mathbf{h}_\i \nabla.\mathbf{u}\\
 +n_\i kT_\i w_{\i s}\frac{\partial u_r}{\partial x_s}+n_\i kT_\i w_{\i t}\frac{\partial u_t}{\partial x_r}+n_\i kT_\i \mathbf{w}_\i \nabla.\mathbf{u}\\
   +\frac{5}{2}n_\i k\mathbf{w}_\i\frac{d T_\i }{dt}+\frac{5}{2} \frac{k}{m_\i}\pi_{\i r}\frac{\partial T_\i}{\partial x_r}+\frac{5}{2} \frac{k}{m_\i}n_\i kT_\i \frac{\partial T_\i}{\partial x_r}=\sum_\j R^{11}_{\i\j}, \label{eq:13n4}
\end{multline}
where $\epsilon_{rst} \text{\CYRZH}_{\i rsl}\omega_{\i t}$ vanishes because $\text{\CYRZH}_{\i rsl}$ is symmetric to the switching of any two indices and the Levi-Civita tensor $\epsilon_{rst}$ is antisymmetric to switching of indices.
Without loss of generality, one may also write the term in the above equation
\begin{multline}
 n_\i kT_\i w_{\i s}\frac{\partial u_r}{\partial x_s}+n_\i kT_\i w_{\i t}\frac{\partial u_t}{\partial x_r}+n_\i kT_\i \mathbf{w}_\i \nabla.\mathbf{u}\\
 =2n_\i kT_\i w_{\i s}\left\{\frac{\partial u_r}{\partial x_s} \right\} +\frac{5}{3}n_\i kT_\i \mathbf{w}_\i \nabla.\mathbf{u}\nonumber,
\end{multline}
as it is found in some literature.

And finally, the balance equation for the stress tensor $n_\i b^{20}_\i=\pi_\i$ is calculated as
 \begin{multline}
 \frac{d\pi_\i}{dt}+\pi_\i\nabla\cdot\mathbf{u}
 +\frac{\partial}{\partial x_r}\left[\text{\CYRZH}_{\i r} +\frac{2}{5}\left(\frac{5}{2}n_\i kT_\i\{ \mathbf{w}_\i\bm{\delta}\}_r +\{\mathbf{h}_\i\bm{\delta}\}_r     \right) \right]\\
 -\left(\frac{X_{\i l}}{m_\i}+\frac{Z_\i eE_{\i l}}{m_\i}+\{\mathbf{u}\times\bm{\omega}_\i\}_l-\frac{d u_l}{dt}\right)\rho_\i\{\mathbf{w}_\i\bm{\delta}\}_l \\
 +\left(\frac{\partial u_r}{\partial x_s}-\epsilon_{rst}\omega_{\i t} \right)\left[ \{\bm{\delta} \pi_{\i s}\}_r+n_\i kT_\i\{\bm{\delta\delta}\}_{rs} \right] =\sum_\j R_{\i\j}^{20} \nonumber
\end{multline}
On manipulating some terms, without loss of generality, and on realizing that $\epsilon_{rst}\{\bm{\delta\delta}\}_{rs}=0$, this takes the form
 \begin{multline}
 \frac{d\pi_\i}{dt}+\pi_\i\nabla\cdot\mathbf{u}+\nabla.\text{\CYRZH}_\i
 +\frac{4}{5}\left\{\frac{\partial}{\partial x_s}\left( \frac{5}{2}n_\i kT_\i {w}_{\i r} +  {h}_{\i r}\right)\right\}\\
 -\rho_\i\left\{{w}_{\i r}\left(\frac{X_{\i s}}{m_\i}+\frac{Z_\i eE_{\i s}}{m_\i}+\{\mathbf{u}\times\bm{\omega}_\i\}_s-\frac{d u_s}{dt}\right)\right\} \\
 +2\left\{\pi_{\i rl}\frac{\partial u_s}{\partial x_l}\right\}+2n_\i kT_\i\left\{\frac{\partial u_r}{\partial x_s}\right\}-2\{\pi_{\i sl}\epsilon_{rst}\omega_{\i t}\}\\
  =\sum_\j R_{\i\j}^{20},\label{eq:13n5}
\end{multline} 
where $\pi_{\i sl}\epsilon_{rst}\omega_{\i t}$ can be thought of as the generalized tensorial cross-product $(\pi_{\i sl}\times\bm{\omega}_{\i})_{rs}$. \footnote{An earlier version of this article contained an error on the first line of Eq.\,(C5). The error is sincerely regretted.}

The equations (\ref{eq:13n1})-(\ref{eq:13n5}) may be considered as the most general moment equations derived from Grad's method in presence of electromagnetic and non-velocity dependent body forces. They agree with Eqs.(4.2.9)-(4.2.14) in Ref.\,\onlinecite{zhdanov_transport_2002}. Notice that the most natural extension of this $13N$-moment scheme would be to a $26N$-moment scheme that incorporates the $1N$ additional moment(s) from $\theta_\i$, $5N$ additional moments from $\sigma_\i$, and $7N$ additional moments from $\text{\CYRZH}_\i$.

It is worth noting that this $13N$-moment system of balance equations is not ``regularized'', meaning that it may not be applicable to cases involving propagation of shocks, doing which may introduce spurious entropy production\cite{struchtrup_regularized_2013,rana_thermodynamically_2016,struchtrup_macroscopic_2005}.

\subsection{General balance equations for moments with $n\geq2$ up to rank-2}
\label{sec:generalbalanceequations}

The general balance equation for rank-0 scalar moments $b^{0n}_\i$, for $n\geq2$ is given by
 \begin{multline}
 \frac{d}{dt}(n_\alpha {b^{0n}_\i})+n_\alpha{b^{0n}_\i}\nabla\cdot\mathbf{u}
 +\frac{\partial}{\partial x_r}\left[n_\i b^{1n}_{\i r} +\frac{n}{\gamma_\i}n_\i b^{1,n-1}_{\i r} \right]\\
 -nn_\i b^{1,n-1}_{\i l}\left(\frac{X_{\i l}}{m_\i}+\frac{Z_\i eE_{\i l}}{m_\i}+\{\mathbf{u}\times\bm{\omega}_\i\}_l-\frac{d u_l}{dt}\right) \\
 +n_\i\left(\frac{\partial u_r}{\partial x_s}-\epsilon_{rst}\omega_{\i t} \right)\left[ nb^{2,n-1}_{\i rs} +\frac{n(n-1)}{\gamma_\i}b^{2,n-2}_{\i rs}+\frac{2n}{3}\bm{\delta}b^{0n}_\i\right.\\
 \left.+\frac{n(2n+1)}{3\gamma_\i}\bm{\delta}b^{0,n-1}_\i\right]
   +\frac{n(2n+1)}{2}\frac{n_\i k}{m_\i}b^{0,n-1}_\i\frac{d T_\i }{dt}\\
   +\frac{n(2n+1)}{2}\frac{n_\i k}{2m_\i}\left[b^{1,n-1}_{\i r}+\frac{n-1}{\gamma_\i}b^{1,n-2}_{\i r} \right]\frac{\partial T_\i}{\partial x_r}\\
   =\sum_\j R^{0n}_{\i\j}. \label{eq:rank0general}
\end{multline}

The general balance equation for rank-$1$ vectorial moments $b^{1n}$, for $n\geq2$ is calculated as
 \begin{multline}
 \frac{d}{dt}(n_\alpha {b^{1n}_\i})+n_\alpha{b^{1n}_\i}\nabla\cdot\mathbf{u} +\frac{\partial}{\partial x_r}\left[n_\i b^{2,n}_{\i r} +\frac{n}{\gamma_\i}n_\i b^{2,n-1}_{\i r}\right.\\
 \left.+\frac{2}{3}\left(\frac{2n+3}{2\gamma_\i}\{n_\i b^{0,n}_\i\bm{\delta}\}_r +\{n_\i b^{0,n+1}_\i\bm{\delta}\}_r     \right) \right]\\
 -\left(\frac{X_{\i l}}{m_\i}+\frac{Z_\i eE_{\i l}}{m_\i}+\{\mathbf{u}\times\bm{\omega}_\i\}_l-\frac{d u_l}{dt}\right)\left[nn_\i b^{2,n-1}_{\i l}\right.\\
 \left.+\frac{2n+3}{3}\{n_\i b^{0,n}_\i\bm{\delta}\}_l \right]+n_\i\left(\frac{\partial u_r}{\partial x_s}-\epsilon_{rst}\omega_{\i t} \right)\left[ nb^{3,n-1}_{\i rs} \right.\\
 \left.+\frac{n(n-1)}{\gamma_\i}b^{3,n-2}_{\i rs}+\frac{2n}{5}\bm{\delta}b^{1n}_\i+\frac{n(2n+3)}{5\gamma_\i}\bm{\delta}b^{1,n-1}_\i+\{\bm{\delta} b^{1n}_{\i s}\}_r\right]\\
   +\frac{n_\i k}{m_\i}\frac{n(2n+3)}{2}b^{1,n-1}_\i\frac{d T_\i }{dt}\\
   +\frac{n_\i k}{m_\i}\frac{n(2n+3)}{2}\left[b^{2,n-1}_{\i r}+\frac{n-1}{\gamma_\i}b^{2,n-2}_{\i r}\right.\\
   \left.+\frac{2}{3}\left(\frac{2n+1}{2\gamma_\i}\{b^{0,n-1}_\i\bm{\delta}\}_r + \{b^{0,n}_\i\bm{\delta}\}_r\right) \right]\frac{\partial T_\i}{\partial x_r}\\
   =\sum_\j R_{\i\j}^{1n},\nonumber 
\end{multline}
which, on manipulating the terms as in the previous subsection,  can be written as\footnote{Line 2 of Eq.\,(C7) contained an error in the earlier version of the manuscript.} 
 \begin{multline}
 \frac{d}{dt}(n_\alpha {b^{1n}_\i})+\frac{2n+5}{5}n_\alpha{b^{1n}_\i}\nabla\cdot\mathbf{u}
 +\frac{\partial}{\partial x_r}\left(n_\i b^{2,n}_{\i r} +\frac{n}{\gamma_\i}n_\i b^{2,n-1}_{\i r}\right)\\
 +\frac{2}{3}\left\{\frac{\partial}{\partial x_r}\left(\frac{2n+3}{2\gamma_\i}n_\i b^{0,n}_\i +n_\i b^{0,n+1}_\i     \right) \right\}\\
 -nn_\i b^{2,n-1}_{\i l}\left(\frac{X_{\i l}}{m_\i}+\frac{Z_\i eE_{\i l}}{m_\i}+\{\mathbf{u}\times\bm{\omega}_\i\}_l-\frac{d u_l}{dt}\right)\\
  -\frac{2n+3}{3}n_\i b^{0,n}_\i\left(\frac{\mathbf{X}_{\i}}{m_\i}+\frac{Z_\i e\mathbf{E}}{m_\i}+\mathbf{u}\times\bm{\omega}_\i-\frac{d \mathbf{u}}{dt}\right) \\
 -n_\i\epsilon_{rst}\omega_{\i t} \bm{\delta}_r b^{1n}_{\i s} +n_\i\left[ nb^{3,n-1}_{\i rs} +\frac{n(n-1)}{\gamma_\i}b^{3,n-2}_{\i rs}\right]\left\{\frac{\partial u_r}{\partial x_s}\right\}\\
 +n_\i\left(\frac{2n+5}{5}b^{1n}_{\i s}\frac{\partial u_r}{\partial x_s}+\frac{2n}{5}b^{1n}_{\i t}\frac{\partial u_t}{\partial x_r}\right)\\
  +\frac{2n(2n+3)}{5}\frac{n_\i}{\gamma_\i}b^{1,n-1}_{\i s}\left\{\frac{\partial u_r}{\partial x_s}\right\}+\left(\frac{n(2n+3)}{5}+\frac{2}{3}\right)\frac{n_\i}{\gamma_\i}b^{1,n-1}_{\i}\nabla\cdot\mathbf{u}\\
   +\frac{n_\i k}{m_\i}\frac{n(2n+3)}{2}b^{1,n-1}_\i\frac{d T_\i }{dt}\\
   +\frac{n_\i k}{m_\i}\frac{n(2n+3)}{2}\left[b^{2,n-1}_{\i r}+\frac{n-1}{\gamma_\i}b^{2,n-2}_{\i r}\right]\frac{\partial T_\i}{\partial x_r}\\
   +\frac{n_\i k}{m_\i}\frac{n(2n+3)}{3}\left(\frac{2n+1}{2\gamma_\i}b^{0,n-1}_\i + b^{0,n}_\i\right) \frac{\partial T_\i}{\partial x_r}\\
   =\sum_\j R_{\i\j}^{1n}. \label{eq:rank1general}
\end{multline}
The general moment equation for rank-$2$ tensorial moments $b^{2n}$, for $n\geq2$, on similar manipulation of terms as for the previous rank-1 moments, is given by
\begin{multline}
 \frac{d}{dt}(n_\alpha {b^{2n}_\i})+\frac{2n+7}{7}n_\alpha{b^{2n}_\i}\nabla\cdot\mathbf{u}
 +\frac{\partial}{\partial x_r}\left(n_\i b^{3,n}_{\i r} +\frac{n}{\gamma_\i}n_\i b^{3,n-1}_{\i r}\right)\\
 +\frac{4}{5}\left\{\frac{\partial}{\partial x_s}\left(\frac{2n+5}{2\gamma_\i}n_\i b^{1,n}_{\i r} +n_\i b^{1,n+1}_{\i r}     \right)\right\}\\
 -nn_\i b^{3,n-1}_{\i l}\left(\frac{X_{\i l}}{m_\i}+\frac{Z_\i eE_{\i l}}{m_\i}+(\mathbf{u}\times\bm{\omega}_\i)_l-\frac{d u_l}{dt}\right)\\
 -\frac{2(2n+5)}{5}\left\{n_\i b^{1,n}_{\i r} \left(\frac{X_{\i s}}{m_\i}+\frac{Z_\i eE_{\i s}}{m_\i}+(\mathbf{u}\times\bm{\omega}_\i)_s-\frac{d u_s}{dt}\right)\right\}\\
 +n_\i\left[ nb^{4,n-1}_{\i rs} +\frac{n(n-1)}{\gamma_\i}b^{4,n-2}_{\i rs}\right]\left\{\frac{\partial u_r}{\partial x_s} \right\}\\
    +n_\i2\left[\frac{2n+7}{7}\left\{b^{2n}_{\i rl}\frac{\partial u_s}{\partial x_l} \right\}+\frac{2n}{7}\left\{b^{2n}_{\i rl}\frac{\partial u_l}{\partial x_s} \right\}\right]\\
     +n_\i\frac{n(2n+5)}{7\gamma_\i}\left[2\left\{b^{2,n-1}_{\i rl}\frac{\partial u_s}{\partial x_l} \right\}+2\left\{b^{2,n-1}_{\i rl}\frac{\partial u_l}{\partial x_s} \right\}+b^{2,n-1}_{\i}\nabla\cdot\mathbf{u}\right]\\
   +n_\i\frac{2(2n+5)}{15}\left(2b^{0,n+1}_\i+\frac{2n+3}{\gamma_\i}b^{0,n}_\i \right) \left\{\frac{\partial u_r}{\partial x_s} \right\}\\
    -2n_\i\{b^{2n}_{\i sl}\epsilon_{rst}\omega_{\i t}\} 
   +\frac{n_\i k}{2m_\i}n(2n+5)b^{2,n-1}_\i\frac{d T_\i }{dt}\\
   +\frac{n_\i k}{m_\i}\frac{n(2n+5)}{2}\left[b^{3,n-1}_{\i r}+\frac{n-1}{\gamma_\i}b^{3,n-2}_{\i r}\right]\frac{\partial T_\i}{\partial x_r}\\
   +\frac{n_\i k}{m_\i}\frac{2n(2n+5)}{5}\left\{\left(\frac{2n+3}{2\gamma_\i}b^{1,n-1}_{\i r} + b^{1,n}_{\i r}\right)\frac{\partial T_\i}{\partial x_s}\right\} \\
   =\sum_\j R^{2n}_{\i\j} \label{eq:rank2general}
\end{multline} 
In general, one can also derive the moment-averaged balance equation for $n<2$ from Eqs.\,(\ref{eq:rank0general})-(\ref{eq:rank2general}), by simply setting the moments with negative indices to zero. \footnote{An earlier version of this article contained errors on lines 2, 4 and 11 of Eq.\,(C8). We would like to thank Jason Hamilton (Cornell University, USA) for pointing out the mistake on line 4.}

\section{The conundrum of $d_{\alpha\beta}$: Why certain values don't work}
\label{sec:dij}

In our previous article, we calculated the collision coefficients of partial bracket integrals in the form suggested by Chapman and Cowling\cite{chapman_mathematical_1952}, Rat et al\cite{rat_transport_2001}. The calculation method in our previous article allowed for a free choice of a factor $d_{\i\j}$, such that on changing $d_{\i\j}$, one could find a new set of collision coefficients. For example, $d_{\i\j}=1$ would represent a calculation similar to the original Chapman and Cowling calculation\cite{chapman_mathematical_1952,bonnefoi_thesis_1975,bonnefoi_thesis_1983,rat_transport_2001}, $d_{\i\j}=\gamma_{\i\j}/2$ would represent the collision coefficients in a form similar to Zhdanov et al\cite{alievskii_1963_transport,zhdanov_transport_2002}, and we chose an additional $d_{\i\j}=\mu_{\i\j}/2kT$ as a reference. The coefficients were then compared to each other, in addition to a single-temperature coefficients, and a range of validity was provided for the the single-temperature coefficients. In this article, we restricted ourselves to only $d_{\i\j}=\gamma_{\i\j}/2$ for the multi-temperature coefficients as referring to only these as ``multi-temperature'' coefficients. We intend to demonstrate the reasoning for this here. There are some steps in the calculation that somewhat obscure a subtlety. The first is the definition of $\Pi^{(m)}_{\i\j}$ of the form 
\begin{multline}
  \Pi_{\i\j}^{(m)}=(1-s)^{-5/2}(1-t)^{-5/2} \pi^{-3} {\mathcal{K}_{\i\j}}\\
  \times \int \{H_{\i\j}^{(m)}(\mathbf{\gn},\chi)-H_{\i\j}^{(m)}(\mathbf{\gn},0)\} g\sigma_{\alpha\beta}(g,\chi)d\Omega d\mathbf{\gn},
\end{multline}
where
\begin{multline}
 H_{\i\j}^{(m)}(\mathbf{\gn},\chi)= \int \exp\left\{  -W_\i^2-W_\j^2-SW_\j^{\prime 2}-TW_\i^2 \right\}\\
 \times P^{(m)}(\mathbf{W}_\j^\prime):P^{(m)}(\mathbf{W}_\i) d\mathbf{X},
\end{multline}
by the absorption of the Sonine polynomials into the exponential. Then this is written as 
\begin{multline}
 H_{\i\j}^{(m)}(\mathbf{\gn},\chi)=\int \exp\left\{ -a_{\i\j}\xn^2-b_{\i\j}\gn^2\right\}\\
 \times P^{(m)}(\mathbf{W}_\j^\prime):P^{(m)}(\mathbf{W}_\i) d\mathbf{\xn}. 
\end{multline}
The exact definitions for the $S$, $T$, $\mathbf{W}$'s, $\mathbf{X}$, $\mathbf{g}$, $a_{\i\j}$, $b_{\i\j}$,   $\mathbf{\xn}$,  $\mathbf{\gn}$ can be found in Appendix A of our previous article\cite{raghunathan_generalized_2021}. For the purposes of this section, it is sufficient to note that $S$  and $T$ are related to the exponential form of the Sonine polynomials, and that $a_{\i\j}$, $b_{\i\j}$,   $\mathbf{\xn}$,  $\mathbf{\gn}$ are expressions of $S$, $T$, $\mathbf{W}$'s, $\mathbf{X}$, $\mathbf{g}$.  The expression is then integrated over $\mathbf{\xn}$, and what remains in the exponential is split and Taylor expanded as follows
\begin{align}
 \exp(-b_{\i\j}\gn^2)&=\exp(-k_{\i\j}\gn^2)\exp((k_{\i\j}-b_{\i\j})\gn^2)\nonumber\\
 &=\exp(-k_{\i\j}\gn^2)\sum_{r=0}^{\infty} \frac{(k_{\i\j}-b_{\i\j})^r}{r!}\gn^{2r},
\end{align}
such that $ \Pi_{\i\j}^{(m)}$ becomes
\begin{equation}
  \Pi_{\i\j}^{(m)}\propto  \sum_{pq\bar{r}\bar{l}} s^p t^q \frac{A^{pq\bar{r}\bar{l},m}_{\i\j}}{k_{\i\j}^{\bar{r}+3/2}}\Omega_{\i\j}^{\bar{l}\bar{r}},
\end{equation}
where the Chapman-Cowling integrals $\Omega_{\i\j}^{\bar{l}\bar{r}}$ are given by
\begin{align}
 \Omega_{\i\j}^{\bar{l}\bar{r}} = &\left(\frac{\pi}{d_{\i\j}}\right)^{1/2}\int_0^\infty \exp(-\zeta^2)   \zeta^{2\bar{r}+3} \phi^{(\bar{l})}_{\i\j} d\zeta,\\
 \phi^{(\bar{l})}_{\i\j}=&\int^\infty_0 (1-\cos^{\bar{l}}{\chi}) \sigma_{\alpha\beta}(g,\chi)\sin{\chi}d\chi,
\end{align}
where $\zeta=k_{\i\j}^{1/2}\gn=d_{\i\j}^{1/2}g$, where 
\begin{equation}
 d_{\i\j}=k_{\i\j}\left\{\mu_{\i\j}^2\left(\frac{\gamma_\i}{2m_\i^2}+\frac{\gamma_\j}{2m_\j^2}\right)\right\},
\end{equation}
which are the $d_{\i\j}$ values that are used in the current article. In the previous article, we had not given any recommendations for the choice of $d_{\i\j}$, only stating that the choice affected the form of the collision coefficients, and they made multi-temperature coefficients significantly different from the single-temperature ones when the temperatures are significantly different. This leads to a few peculiarities. For example, as long as the value of $d_{\i\j}$ are positive and only depend on the masses and temperatures of the colliding species, one can have a multitude of different collision coefficients which in principle should be the same. However, once the collision potential is chosen, they provide values of collision coefficients which are different for different $d_{\i\j}$, seemingly making the solution multi-valued. Furthermore, in the current article, we find spurious singularities in the collision coefficients calculated from certain choices of $d_{\i\j}$.

Now, with $d_{\i\j}$, the overall exponential term in $H_{\i\j}^{(m)}$, on integrating over $\mathbf{\xn}$ can be written as
\begin{multline}
 H_{\i\j}^{(m)}(\mathbf{\gn},\chi)\propto \exp\left\{ -d_{\i\j}g^2\right\}R_m(S,T,\gn,\cos{\chi})\\
 \times  \sum_{r=0}^{\infty}\frac{(d_{\i\j}g^2-b_{\i\j}\gn^{2})^r}{a_{\i\j}^{m+3/2}r!}, 
\end{multline}
where $R_m$ is some scalar function of $\gn$ and $\cos{\chi}$ depending on the rank $m$ of the tensor. It can be noted that, at this point this expresses $H_{\i\j}^{(m)}$ as a convergent infinite series. However, the rate of convergence has not been addressed in the previous literature, as the series has not been expressed in this explicit form. This is the subtlety alluded to at the beginning of this note. 

It is difficult to directly judge the rate of convergence of the series from this expression since it still depends on $(S,T)$ through $b_{\i\j}$. Therefore, it is instructive to try to see what this series aims to compute. We can define an $\mathcal{H}_{\i\j}^{(m,p,q)}$ of the following form
\begin{multline}
 \mathcal{H}_{\i\j}^{(m,p,q)}= \int \exp\left\{  -W_\i^2-W_\j^2 \right\}S^p_{m+1/2}(W_\j^{\prime2})S^q_{m+1/2}(W_\i^2)\\
 \times P^{(m)}(\mathbf{W}_\j^\prime):P^{(m)}(\mathbf{W}_\i) d\mathbf{X}.
\end{multline}
It is straightforward to notice that the expressions $H^{(m)}_{\i\j}$ should reduce to $\mathcal{H}^{(m,p,q)}_{\i\j}$ for any $(p,q)$. Now, from the relations between the $\mathbf{W}$'s, $\mathbf{X}$ and $\mathbf{\gn}$, it is straightforward to write this as 
\begin{multline}
  \mathcal{H}_{\i\j}^{(m,p,q)}\propto \int \exp\left\{  -\xn^{\prime2}-b_{\i\j}^{\prime}\gn^{2} \right\}S^p_{m+1/2}(W_\j^{\prime2})S^q_{m+1/2}(W_\i^2)\\
 \times ({W}_\j^\prime{W}_\i)^m P_m(\mathbf{\hat{W}}_\j^\prime.\mathbf{\hat{W}}_\i) d\mathbf{\xn}.
\end{multline}
where $\mathbf{\xn}^\prime=\mathbf{X}-{(M_{\i1}M_{\i2})^{1/2}}(1-\theta_{\i\j})\mathbf{\gn}$, and $b_{\i\j}^{\prime}=1-M_{\i1}M_{\i2}(1-\theta_{\i\j})^2$, and where $P_m$ is the Legendre polynomial. Now, if the Sonine and Legendre polynomials are series expanded, and then the $W$'s are expressed in terms of $\xn^{\prime2}$ and $\gn^{(\prime)2}$, and integrate over $4\pi\xn^2 d\xn$, we will find that $\mathcal{H}_{\i\j}^{(m,p,q)}$ evaluates to a finite series in $\gn^2$ and $\cos{\chi}$. The maximum power of $\gn^2$ in this series is $p+q+m$. 

Now, one can notice immediately that $b_{\i\j}^{\prime}\gn^2=[1-M_{\i1}M_{\i2}(1-\theta_{\i\j})^2]\gn^2=(\gamma_{\i\j}/2) g^2$. Therefore, in the calculation of the bracket integrals in our previous article, for $d_{\i\j}=\gamma_{\i\j}/2$, the series for $H^{(m)}_{\i\j}$ converges perfectly on just retaining $r\leq p+q$ terms in the summation, with additional terms vanishing (because $R_m$ usually has a power of $\gn^{2m}$). 

It also implies that other choices of $d_{\i\j}$ are not fully converged on retaining any finite number of terms, because in $\mathcal{H}_{\i\j}^{(m,p,q)}$ with a different choice of $d_{\i\j}$, we would have
\begin{multline}
  \mathcal{H}_{\i\j}^{(m,p,q)}\propto \int \exp\left\{  -\xn^{\prime2}-k_{\i\j}\gn^{2} \right\}\\
  \times S^p_{m+1/2}(W_\j^{\prime2})S^q_{m+1/2}(W_\i^2)({W}_\j^\prime{W}_\i)^m P_m(\mathbf{\hat{W}}_\j^\prime.\mathbf{\hat{W}}_\i)\\
  \times \sum_{r^\prime}^{\infty}\frac{(b_{\i\j}^{\prime}-k_{\i\j})^{r^{\prime}}}{r^{\prime}!}\gn^{2r^{\prime}} d\mathbf{\xn},
\end{multline}
which turns $\mathcal{H}_{\i\j}^{(m,p,q)}$ into an infinite series. In our previous article, we had recommended a general limit of $r\leq p+q$ based on a comparison of terms expressed in terms of Chapman-Cowling integral and the series sum in ${H}_{\i\j}^{(m)}$ as was done in Refs.\,\onlinecite{chapman_mathematical_1952,bonnefoi_thesis_1975,rat_transport_2001}. This explains why we do not observe singularities with the choice of $d_{\i\j}=\gamma_{\i\j}/2$, and that the singularities with other values of $d_{\i\j}$ are a result of truncating the series too early.
In Refs.\,\onlinecite{chapman_mathematical_1952} and \onlinecite{bonnefoi_thesis_1975}, this leads to the correct result as $b_{\i\j}^{\prime}=k_{\i\j}$ when the temperatures are equal, thus retaining only the $r^{\prime}=0$ term. Thus, we take the opportunity to state that for the multi-temperature bracket integrals, only $d_{\i\j}=\gamma_{\i\j}/2$ will lead to a finite converged series for ${H}_{\i\j}^{(m)}$ with $r\leq p+q$, and the rest will remain infinite sums as indicated in the expression. It may be possible to remedy the singularities by adding more terms to the series in the calculation of the bracket integral. Since however, a converged series is already available for $d_{\i\j}=\gamma_{\i\j}/2$, to do so would be a quixotic affair.  Thus, the values of the collision coefficients generate with $d_{\i\j}=\gamma_{\i\j}/2$  may be considered the ``closed forms'' for the multi-temperature bracket integrals. Based on this, the generalization of the bracket integral calculation to an arbitrary rank-$m$ is going to comprise a part of our future work.

\end{document}